\documentclass[11pt, a4paper,preprintnumbers]{article}

\usepackage{jheppub}
\usepackage{amsmath}
\usepackage{graphicx,amsmath,amssymb}
\usepackage{subcaption}
\usepackage{caption}
\usepackage{tabularx}
\preprint{LAPTH-030/24}

\title{On the number of Regge trajectories for dual amplitudes}

\author[a,b]{Christopher Eckner,} 
\author[b]{Felipe Figueroa,}
\author[b]{Piotr Tourkine}

\affiliation[a]{Center for Astrophysics and Cosmology, University of Nova Gorica, Vipavska 11c, 5270 Ajdov\v{s}\v{c}ina,
Slovenia}
\affiliation[b]{LAPTh, CNRS et Universit\'{e} Savoie Mont-Blanc 
9 Chemin de Bellevue, F-74941 Annecy, France}

\abstract{Regge poles connect the analytic structure of scattering amplitudes, analytically continued in angular momentum, to their high-energy limit in momentum space. Dual models are expected to have only Regge poles as singularities in angular momentum space, and string theory suggests there should be an infinite number of them. In this study, we investigate the number of Regge trajectories these models may have. We prove, based solely on crossing symmetry and unitarity, that meromorphic amplitudes, with or without subtractions, cannot produce a reggeizing amplitude if they contain any finite number of Regge trajectories, and show that this excludes the existence of such amplitudes altogether. Additionally, we develop and apply a linear programming dual bootstrap method to exclude these amplitudes directly in momentum space. 
}

\begin{document}
\maketitle

\section{Introduction}

One of the goals of the modern $S$-matrix bootstrap~\cite{Paulos:2016fap,Paulos:2016but,Paulos:2017fhb,Kruczenski:2022lot} is to characterize the space of $S$-matrices consistent with the principles of unitarity, analyticity and crossing symmetry.

Within this space, dual model scattering amplitudes constitute a subset of $S$-matrices that describe the exchange of a tower of higher spin resonances at tree-level, namely they are cases of weakly interacting theories of massive higher spins~\cite{Caron-Huot:2016icg}. Because they are purely meromorphic, they do not satisfy full non-perturbative unitarity, but only positivity. Nevertheless, their simplicity renders them some of the simplest scattering amplitudes in the space of $S$-matrices, and thus ideal building blocks for further exploration.

Tree-level open string theory is the prime example of a dual model, but others exist and recent years have seen a renewal of activity on this question, with the discovery of a plethora of new models\footnote{See for instance \cite{Haring:2023zwu,Cheung:2023adk,Cheung:2023uwn,Geiser:2022exp}.}. Large-$N$ gauge theories are also weakly coupled and should also be described by  meromorphic scattering amplitudes~\cite{tHooft:1973alw,Witten:1979kh}, and have seen remarkable progress recently~\cite{Albert:2022oes,Albert:2023jtd,Albert:2023seb,Fernandez:2022kzi,Ma:2023vgc}.

Dual models also exhibit one salient feature of non-perturbative behaviour: reggeization. At high energies $\sqrt{s}$ and fixed momentum transfer $\sqrt{-t}$, reggeizing amplitudes behave as $A(s,t)\sim s^{\alpha(t)}$. The Regge trajectory function $\alpha(t)$ also determines the spectrum of the tower of resonances $m_0^2,m_1^2,\dots$ of a given theory, and thus reggeization relates the spectrum to high-energy scattering in a profound manner. In various instances of recent studies of non-perturbative $S$-matrices, intriguing Regge trajectories emerge~\cite{Guerrieri:2023qbg,Guerrieri:2021ivu,Guerrieri:2022sod,Acanfora:2023axz,Gumus:2023xbs,EliasMiro:2022xaa}. Gaining a deeper understanding of Regge theory is a timely question. From this perspective, dual models provide an excellent testing ground. 

In this context, the authors of the present paper recently developed a numerical bootstrap approach to explore the space of dual model amplitudes~\cite{Eckner:2024ggx}. Little is known of this space because the constraint of channel duality -- that is, the poles in one channel are reproduced exactly by the summation in the cross-channel -- is particularly hard to satisfy.

In this paper, we explore an aspect of this question: 
\emph{How simple can dual models be?} 
Arguably, the simplest dual model would be composed of exactly one Regge trajectory; it would make an elegant object, parameterized by its spectrum composed of particles of masses $m^2_n\big|_{n\in\mathbb{N}}$ and spin $n$. The amplitude would read:
\begin{equation}
  A(s,t)=  -\sum_{n=0}^\infty c_n \frac{P_n\!\left(1+\frac{2t}{\vphantom{\intop^2}m_{n}^2-4\mu^2}\right)}{\vphantom{\intop}s-m_n^2}
  \label{eq:dm-srt}
\end{equation}
where $P_n$ are Legendre polynomials. 
While the details of this expression will be explained later, the important point to notice is that in contrast to tree-level string theory for instance, at each level $n$, there is only \textit{one} state of spin $n$, instead of a sum over spins $J=0,\dots,n$. Thus, there is only one unknown per level, the residue $c_n$, required to be positive by the condition of unitarity $c_n\geq0$. Therefore, such amplitudes would be parameterized by a \textit{linearly growing} number of parameters; this is to be compared to the quadratic number of parameters seen in string theory, and thus could offer a numerically efficient way to study dual models, provided such objects existed. 

In addition to this simplicity, the question of the existence of meromorphic amplitudes with finitely many trajectories is timely given the aforementioned works~\cite{Albert:2023seb,Albert:2022oes,Albert:2023jtd} where, through an elaborate bootstrap procedure, the authors have identified a special amplitude argued to correspond to pion-pion scattering in large $N$ QCD. In this case, the leading trajectory significantly dominates the sub-leading ones, raising the question of whether it could consist of a single trajectory.

In this paper, we prove that meromorphic amplitudes of external scalars with finitely many Regge trajectories and with zero or one subtraction are inconsistent with the standard assumptions of unitarity, analyticity and crossing symmetry.  The notion of subtractions refers to the behaviour at infinity and is described below; for us it is essential to discuss once-subtracted amplitudes in order to connect with~\cite{Albert:2023seb,Albert:2022oes,Albert:2023jtd}. 

The proof is rather simple and relies on well-known ingredients of the $S$-matrix machinery: the Froissart-Gribov projection, analytic continuation of the angular momentum into complex values, and the use of crossing symmetry. We show that singularities in the crossed channel dictate some particular decay for the coefficients of the pole expansion of the amplitude. Then, using the Froissart-Gribov projection, we can show that this decay is too strong to allow the partial wave coefficients to exhibit a Regge pole in the angular momentum, and conclude that the amplitudes do not reggeize. Finally, we show that the absence of Regge poles clashes with having a well-defined partial wave expansion for the amplitude, with the partial wave coefficients diverging for physical values of the energy. This leads us to conclude that meromorphic amplitudes containing finitely many trajectories are inconsistent objects, and therefore cannot exist.

{\it En passant}, we review a construction known as the Mandelstam-Sommerfeld-Watson transform~\cite{Mandelstam:1963iyb} that, while not strictly necessary for proving the aforementioned results, constitutes a useful yet somewhat unnoticed tool, which we present in a modern language for the benefit of the bootstrap community. Moreover, based on this construction we conjecture the possibility of a complex angular momentum space version of channel-duality, the defining feature of dual model amplitudes given by the fact that they can be described in terms of infinite sums of singularities in only one channel, with the singularities in the crossed channel emerging from the divergence of this infinite series.

We then provide an argument based on a construction known as the Mandelstam-Sommerfeld-Watson transform~\cite{Mandelstam:1963iyb} that permits, in principle, to convert the absence of Regge poles into a statement of non-existence of amplitudes in momentum space. We discuss the assumptions and possible caveats of this argument.

Finally, as an alternative way of arriving at the results mentioned above, we develop a numerical dual bootstrap to study meromorphic amplitudes directly in momentum space. We use it to study the leading trajectory of~\cite{Albert:2023seb,Albert:2022oes,Albert:2023jtd} and show that such amplitudes cannot be made of a single Regge trajectory.

We also discuss where the presence of infinitely many trajectories lets us bypass the non-existence argument. In particular, we observe that our argument yields no constraint on the size of the sub-leading trajectories: they can be as small as desired, as long as they come in an infinite number. This statement, therefore, suggests that the large $N$ amplitude of~\cite{Albert:2023seb,Albert:2022oes,Albert:2023jtd} should have an infinite number of trajectories, but they might require extra numerical efforts to be detected. It would be interesting to study this point further.

\paragraph{Remark.} 
After we had completed the proof presented in this paper, we discovered that a few references had already considered the question of the number of Regge trajectories during the 60s. We discuss them in section.~\ref{sec:discussion}. 
Some of them
rely on the use of the so-called Finite Energy Sum Rules (FESRs), which involve making a certain number of assumptions which we discuss in appendix~\ref{app:fesr}. Our initial attempts to address the problem also relied on a version of FESRs, which we had independently re-discovered. However, we were unsatisfied with the assumptions going into their applicability. The argument presented in this chapter allows us to avoid any such assumptions and also rules out the in principle possible case in which the dual model amplitude is not made up of Regge poles, but rather by infinitely many isolated poles in the Mandelstam variables\footnote{Isolated in the sense that they are not connected by a trajectory function $\alpha(t)$ that interpolates between them.}. Most importantly, we also work out the case with subtractions, which was not handled by these earlier papers, and which is relevant for the aforementioned pion amplitude at large~$N$ in particular, but also for the $S$-matrix program in general.%

Another approach claimed to identify spurious singularities in the case of unequal masses for the external scattered states, which required infinitely many extra trajectories to cancel them out~\cite{Freedman:1966zec,Nachtmann:2003ik}. We review the argument in the discussion section, and we comment that the proof relies on an ingredient that appears essentially impossible to justify without assuming extra conditions on the behaviour of the analytically continued partial wave coefficients in complex angular momentum. Therefore, this approach is inconclusive as it stands.\footnote{We thank A. Sinha and A. Vicchi for pointing us to this reference, as well as the organizers of the 2024 S-matrix bootstrap for putting together the conditions for high-quality scientific interactions to take place.}

\paragraph{Plan of the paper} The paper is organized as follows. In section~\ref{sec:setup}, we review some basic elements of $S$-matrix and Regge theory which we need later, in particular, the Froissart-Gribov projection and the Sommerfeld-Watson transform, as well as its improved version by Mandelstam. In section~\ref{sec:ruling-out}, we present the proof that dual model amplitudes with finitely many trajectories are inconsistent. In section~\ref{sec:linear-programming}, we introduce a linear-programming dual bootstrap and apply it to the power-law leading Regge trajectory of~\cite{Albert:2023seb,Albert:2022oes,Albert:2023jtd}. In section~\ref{sec:discussion}, we discuss in detail the previous literature on Regge trajectories, summarize our main arguments, and mention some future directions.

\section{Dual models and Regge poles: analyticity, crossing, unitarity}
\label{sec:setup}
In this section, we present the analytic structure of the scattering amplitudes we work with, and elaborate on their properties of unitarity and crossing symmetry. We study the case of identical external scalars of mass $\mu^2$ with colour-ordering.

\subsection{Crossing symmetry}
We assume crossing symmetry of our scattering amplitude. Because of the color-ordering, instead of having full $s$-$t$-$u$ crossing symmetry, the amplitude is only invariant under the exchange of $s$ and $t$, and hence crossing symmetry is simply:
\begin{equation}
    A(s,t)=A(t,s)\,.
\end{equation}

\subsection{Analyticity and channel duality.}
\begin{figure}
    \centering
    \includegraphics{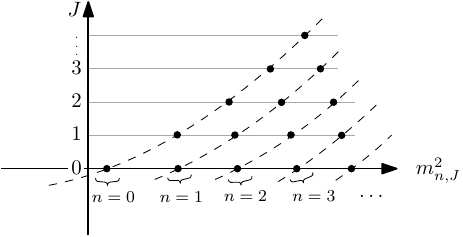}
    \caption{Typical spectrum in a dual model amplitude, organized by Regge trajectories (dashes) and where the levels $n=0,1,2,\dots$ are explicitly shown. Note that with non-linear trajectory, integer spacing between trajectories (as in string theory) generically yields a non-degenerate spectrum of masses, i.e., within the same level $n$ all masses $m^2_{n,J}$ are distinct.}
    \label{fig:dual-model}
\end{figure}

As stated in the introduction, we work with scattering amplitudes that are purely meromorphic. They describe the exchange of higher spin particles of masses $m_{n,J}^2$ at tree-level. They can be organized in Regge trajectories, or levels $n$, as shown in fig.~\ref{fig:dual-model}.

With such a spectrum $m_{n,J}^2$, a simple dispersion relation yields the following representation of the amplitude:
\begin{equation}
    A(s,t) = \frac{1}{2i\pi}\oint \frac{A(s',t)}{s'-s}\,\mathrm{d}s' = -\sum_{n,J} \frac{{\rm Res} A(s^{\prime},t)|_{s^{\prime}=m_{n,J}^2}}{m_{n,J}^2-s}+A_\infty(t)
\end{equation}
where $A_\infty(t)$ represent the contribution of the contour at infinity, following the notation of~\cite{Cheung:2023uwn}. When a spin-$J$ state is being exchanged at $s=m_{n,J}^2$, the residue at the pole as a function of $t$ is given by a Legendre polynomial $P_n(\cos(\theta_{n,J}))$, where $\cos(\theta_{n,J}) = 1+\frac{\vphantom{\intop^2}2t}{\vphantom{\intop^2}m_{n,J}^2-4 \mu^2}$:
\begin{equation}
    \textrm{Res}\, A(s',t)|_{s'=m_{n,J}^2} =c_{n,J} P_J\!\left(1+\tfrac{\vphantom{\intop^2}2t}{\vphantom{\intop^2}m_{n,J}^2-4 \mu^2}\right), 
\end{equation}
up to an undetermined proportionality constant $c_{n,J}$ that corresponds to the square of the coupling between the external scalar particle and the exchanged spin-$J$ state. The ultimate reason for this is the textbook material fact that there exists a unique interaction vertex between a spin-$J$ and two scalar particles that is fixed by Lorentz invariance.\footnote{The vertex is simply made up of anti-symmetrized derivatives of the scalar field, contracted with the spin-$J$ indices,  see for instance~\cite{Caron-Huot:2016icg,Arkani-Hamed:2020blm} and references therein.}

\textit{If} the amplitude decays at infinity in $s$, the arc drops and we obtain the representation of the amplitude
\begin{equation}
\label{eq:dualityunsubtracted}
    A(s,t) =  \sum_{n,J} c_{n,J} \frac{P_J\!\left(1+\tfrac{\vphantom{\intop^2}2t}{\vphantom{\intop^2}m_{n,J}^2-4\mu^2}\right)}{\vphantom{\intop^2}s-m_{n,J}^2}.
\end{equation}
This expression of the amplitude was derived with the sole assumption that it decays at infinity. Therefore, crossing implies that poles in $t$ hide in this expression. In particular, we expect that the series should diverge when $t$ hits the first singularity, $t=m_{0}^2$ in such a way that $A(s,t)\sim_{t\to m_0^2} \frac1{t-m_0^2}$. {Note that we are not assuming that the mass of the external particle $\mu^2$ is necessarily equal to $m_0^2$}.

This property called channel duality is simply a consequence of the decay of the amplitude at infinity, namely the fact that it satisfies unsubtracted dispersion relations in $S$-matrix theory jargon.\footnote{The name was coined in~\cite{Dolen:1967jr}.}

If the amplitude does not decay at infinity, we need to proceed to a ``subtraction'' procedure to write down such dispersion relations. Assuming that the amplitude is bounded by $s$ at large energies, we perform the same Cauchy contour deformation as above but for the function $A(s,t)/s$:
\begin{equation}
\label{eq:DR-1sub-cauchy}
    \frac{A(s,t)}s =\frac{1}{2i \pi} \oint_{|s'-s|=\epsilon} \frac{A(s',t)}{s'(s'-s)}\,\mathrm{d}s' =\frac{A(0,t)}{s} + \sum_{n,J} c_{n,J}\frac{P_J\!\left(1+\tfrac{\vphantom{\intop^2}2t}{\vphantom{\intop^2}m_{n,J}^2-4\mu^2}\right)}{\vphantom{\intop^2}m_{n,J}^2(s-{m_{n,J}^2)}}.
\end{equation}
At this point, we can simplify the expression by setting $t$ to zero on the left-hand side:
\begin{equation}
    A(s,0) = A(0,0) + \sum_{n,J} c_{n,J}\frac{s}{\vphantom{\intop^2}m_{n,J}^2(s-{m_{n,J}^2)}}
\end{equation}
where we used that for all $J$, $P_J(1)=1$. 
By virtue of crossing, $A(s,0)=A(0,s)$. Thus, the expression above tells us what $A(0,t)$ is in eq.~\eqref{eq:DR-1sub-cauchy} in terms of the $c_{n,J}$'s and an undetermined constant $A(0,0)$, so that we arrive at the final dispersion relation for the amplitude:
\begin{equation}
\label{eq:DR-1sub}
    A(s,t) = A(0,0)+\sum_{n,j} \frac{c_{n,J}}{m_{n,J}^2}\left( \frac{s\, P_J\!\left(1+\tfrac{\vphantom{\intop^2}2t}{\vphantom{\intop^2}m_{n,J}^2-4\mu^2}\right)}{s-{m_{n,J}^2}} + \frac{t}{t-{m_{n,J}^2}} \right).
\end{equation}

Compared to the case of unsubtracted duality, we can see that both, the singularities in $s$ and $t$, appear summed independently. Note that the residues at $s$-singularities can be read off the first sum directly, as when $s\to m_{n,J}^2$, the factor of $s$ in the numerator and the $1/m_{n,J}^2$ exactly cancel out each other to produce the correctly normalized spin-$J$ exchange term. However, the residue of the  $t$ channel singularities is incorrect (except for $n=J=0$), which means that the summation over the $s$ channel singularities must continue to provide the missing bits that will produce a full Legendre polynomial, and hence the summation must also cease to be convergent at these singularities just like for the unsubtracted case.

\subsection{Unitarity}
Dual models correspond to weakly coupled theories, and thus they are not restricted by the full non-perturbative unitarity constraints in the form of the optical theorem. More modestly, in the case of dual models unitarity, boils down to the requirement that the residue coefficients $c_{n,J}$ must be non-negative:
\begin{equation}
    c_{n,J}\geq0\rm{,}
\end{equation}
which reflects the fact that on a pole, the four-point amplitude factorizes as the product of three-point vertices $c_{n,J} \sim \lambda_{J,\phi,\phi}^2$ where $\lambda_{J,\phi,\phi}$ is the coupling of the spin-$J$-scalar-scalar vertex. For the Lagrangian to be hermitian, this coupling needs to be real, which yields the positivity above.

\subsection{Partial wave decomposition; Froissart-Gribov projection}
We now turn to the study of partial waves. We explain first the Froissart-Gribov projection, which allows one to continue the partial waves at complex spin. Then we describe the analytic structure of the partial waves in the complexified angular momentum plane. We refer for instance to the textbooks~\cite{Collins_1977,Gribov:2003nw}, and the recent account~\cite{Correia:2020xtr}.

While in quantum mechanics and quantum field theory, the angular momentum is of course quantized, the Froissart-Gribov formula reviewed below allows one to analytically continue the partial waves into complex values of the angular momentum $J$, so that one can effectively trade the continuous momentum transfer squared $\sqrt{-t}$ for a continuous spin~$J$: $$A(s,t)\mapsto A_J(s)\,.$$

Note that in this section and the paper, we work in $d=4$ for notational convenience, but our results will extend to higher dimensions. 
\paragraph{Partial wave expansion.}
We start with the partial wave expansion in the physical $s$ channel, $s>4\mu^2,\,t<0$:
\begin{equation}
    A(s,t)=16 \pi \sum_{J=0}^\infty (2J+1) A_J (s) P_J (z_s)
    \label{eq:PWE}
\end{equation}
where $z_s$ is the cosine of the scattering angle in the $s$ channel,
\begin{equation}
\label{eq:costheta}
z_s = 1+\frac{2t}{s-4\mu^2}
\end{equation}
and $P_J$ are Legendre polynomials.

The partial-wave sum runs over all spins\footnote{Odd spins would drop out if we relaxed the "colour-ordering" and had full $s-t-u$ symmetry}. It converges for all physical $s$. In the $t$ plane it diverges at the first singularity, $t=4\mu^2$ when there are no bound states, or at the first bound state  $t=m_0^2$. The Legendre polynomials satisfy the following orthogonality relation:
\begin{equation}
    \left(J+\frac 12\right)
    \int_{-1}^1 P_J(z)
    P_{K}(z)
    \,\mathrm{d}z = \delta_{J,K},
\end{equation}
which allows one to invert the sum and define the partial wave projection:
\begin{equation}
    f_J(s) =
    \frac1{16\pi}
    \frac{1}{s-4\mu^2}\int_{4\mu^2-s}^0  P_J(1+\frac{2t}{s-4\mu^2}) A(s,t)\,\mathrm{d}t\rm{,}
\end{equation}
where we switched back to an integral over $t$ using eq.~\eqref{eq:costheta}.

\paragraph{Froissart-Gribov}
As mentioned before, the partial wave projection above diverges when $t$ hits the first singularity $t=t_0$, which restricts the possibility of using crossing symmetry on the partial wave expansion of the amplitude. 

\begin{figure}
    \centering
    \includegraphics[scale=1.2]{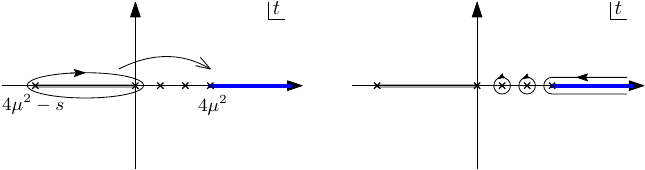}
    \caption{Froissart-Gribov contour deformation in the presence of bound states below the threshold. In our case, we have no threshold cut but only an infinite sequence of poles.}
    \label{fig:FG}
\end{figure}

The Froissart-Gribov projection allows one to bypass this problem, and also to analytically continue the partial waves into complex spin. The main idea of the derivation is to use the fact that the Legendre $P$ polynomials are the discontinuity of the Legendre $Q$ functions, defined for instance in~\cite{Correia:2020xtr} whose conventions we follow.
For integer $J\in \mathbb{Z}$, we have
\begin{equation}
{\rm Disc}_z \left(Q_J(z)\right)=-\frac{\pi}2P_J(z)\,,\quad z\in[-1;1]
\end{equation}
where $\rm Disc$ is defined as 
\begin{equation}
   {\rm{Disc}}_z f(z):=\frac1{2i} (f(z+i\epsilon)-f(z-i \epsilon))\,.
\end{equation}
Using this, the line integral in the partial wave transform can be turned into a contour integral encircling $[-1;1]$. This contour can then be extended to infinity and winds around the cut starting at the first singularity in the $z$-plane, as shown in fig.~\ref{fig:FG}. The condition under which the arc at infinity can be dropped depends on the spin $J$, and on the fall-off of the amplitude: For amplitudes that reggeize, we expect that they behave as $s^{\alpha(t)}$ at large $s$, fixed $st$, and since the $Q$-functions behave as
\begin{equation}
\label{eq:QJasymptotic}
    Q_J(z)\sim_{z\to\infty} \frac1{z^{J+1}},
\end{equation}
we see that the arc will drop if $J>\alpha(t)$. 

In the case of once-subtracted dispersion relations, we know that the amplitude might approach a constant at infinity, and we obtain analyticity in $J$ for $J>\rm{max}[\,0,\alpha(t)\,]$.

Putting all these elements together we arrive at the Froissart-Gribov projection:
\begin{equation}
\label{eq:FGformula}
 f_J(s) ={1\over16 \pi^2}\frac{2}{s-4\mu^2}\int\limits_{t_0}^\infty \mathrm{d}z \,Q_J\left(1+\frac{2t}{s-4\mu^2}\right){\rm Disc}_t A(s,t)\,, 
\end{equation}
where $t_0$ is the location of the first singularity of the amplitude in the $t-$channel.

Crucially    for us, this formula can be specialized in the case of meromorphic amplitudes, where all the singularities are simple poles. In this case, we get a discrete sum over residues:
\begin{equation}
\label{eq:FGformula-poles}
  f_J(s) ={1\over16 \pi^2}\frac{2}{s-4\mu^2} \sum_{n=0}^\infty {\oint}_{|t-t_n|=\epsilon} \mathrm{d}t \,Q_J\left(1+\frac{2t}{s-4\mu^2}\right) A(s,t)\,, 
\end{equation}

The residues of the amplitude at $t=m_{n,J'}^2$ are Legendre polynomials in $s$, as explained above. When $t=m_{n,J'}^2$, we have that $z=z_{n,J'}=1+\frac{2m_{n,J'}^2}{s-4\mu^2}$, and finally
\begin{equation}
\label{eq:FGformula-poles-residues}
    f_J(s) ={i\over4 \pi}\frac{1}{s-4\mu^2}  \sum_{n=0}^\infty\sum_{J'=0}^n c_{n,J'}\,
    Q_J\!\left(1+\frac{2m_{n,J'}^2}{s-4\mu^2}\right) P_{J'}\!\left(1+\frac{2s}{m_{n,J'}^2-4\mu^2}\right)\,.
\end{equation}

\subsection{Regge poles}

Having shown that the partial wave amplitudes can be analytically continued in the complex $J$-plane, we can now answer the question that relates the analytic structure in the $J$ to the high-energy behaviour of the amplitude.

\begin{figure}
    \centering
    \includegraphics{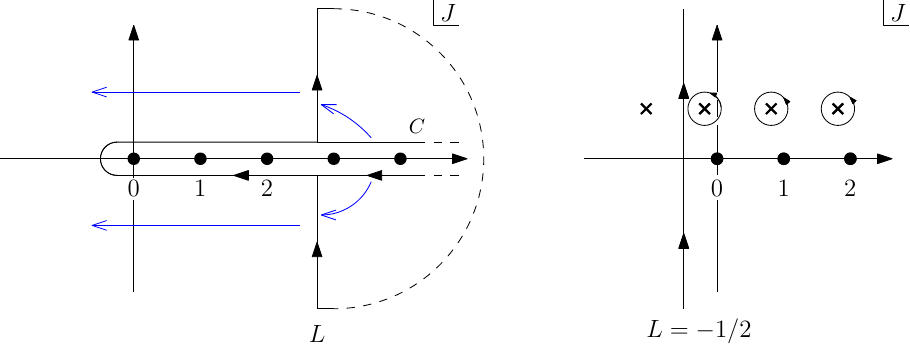}  
    \caption{Sommerfeld-Watson transform. First turn the sum over discrete spins $J$ as a continuous integral, contour $C$. Then open up the contour (blue arrows) to obtain a vertical integral. The vertical integral can be pushed to $L=1/2$ where the Legendre $P$ functions are the softest, which decreases the magnitude of the resulting vertical integral, called the background integral.}
    \label{fig:SW}
\end{figure}

Equipped with a definition of the partial wave at complex values of $J$, we can use a tool called the Sommerfeld-Watson transform to replace the discrete sum over $J$ in \eqref{eq:PWE} with an integral along a contour $C$ shown in fig.~\ref{fig:SW} as follows:
\begin{equation}
\label{eq:SommerfeldWatson}
    A(s,t)=-\frac{16 \pi}{2i} \int_C \frac{\mathrm{d}J}{\sin \pi J} (2J+1) f_J (t) P_J (-z_t)
\end{equation}
where the sine function in the denominator picks up the residues at non-negative integer values of $J$, and where we have used that $\oint_{|J-n|=\epsilon} \frac{\mathrm{d}J}{\sin(\pi J)}=2i (-1)^n$ and that $P_J(-z) = (-1)^J P_J(z)$ for integer $J$.

This integral representation of the amplitude allows us to deform the contour along the imaginary axis, as shown in fig.~\ref{fig:SW}. The asymptotics of the Legendre $P$ functions is given by
\begin{equation}
\begin{aligned}
    &P_J(z) = e^{i J \theta}+e^{-i J \theta}\\
    &P_J(-z) = e^{i J (\theta-\pi)}+e^{-i J (\theta-\pi)}
\end{aligned}
\end{equation}
where $z=\cos(\theta)$. Thus, in the physical region $-1<\cos(\theta)<1$, the factor $P_J(-z)/\sin(\pi J)$ decays exponentially as $\text{Im} \,J \to \infty$ and lets us drop the arc at infinity, since the partial waves themselves decay exponentially on the right half $J$-plane as a consequence of the Froissart-Gribov projection.

As the contour is deformed it picks up possible singularities in $J$ of the integrand,
and once the contour deformation is done the expansion is defined on the entire $t$-plane (minus the cuts), far beyond its initial range of convergence which was given by the nearest singularity in $t$.

The amplitudes we construct are meromorphic. Hence, the only possible singularities in $J$ are simple poles located at some positions $J=\alpha(t)$. Otherwise, branch-cut singularities in $J$ would introduce non-meromorphy in the Mandelstam variables. These types of partial wave singularities at analytically continued momenta are called \textit{Regge poles}, and their residues are  given by a function $\beta(s)$, so that near a pole $J=\alpha(s)$, $s$ fixed, we have
\begin{equation}
    f_J(s) \simeq \frac{\beta(s)}{J-\alpha(s)}\rm{.}
\end{equation}
In general, there can also be cuts when loops are included, see e.g. the textbook~\cite{Gribov:2003nw}. 

Overall we obtain that in the case of meromorphic amplitudes,
\begin{equation}
\label{eq:WS-RP}
    A(s,t) = -\frac{16\pi}{2i}\int_{-1/2-i\infty}^{-1/2+i \infty}\frac{\mathrm{d}J}{\sin \pi J} (2J+1) f_J (t) P_J (-z_t) +A_{\rm Regge-poles}(s,t),
\end{equation}
where
\begin{equation}
A_{\rm Regge-poles}(s,t) = -16\pi^2\sum_{i}(2\alpha_i(t)+1)\beta_i(t) \frac{P_{\alpha_i(t)}(-z_t)}{\sin(\pi \alpha_i(t))}
\end{equation}

In the r.-h.-s. of \eqref{eq:WS-RP} the integral is called the \textit{background} integral. 
The choice to set the contour of the background integral at $\text{Re} \,J = -1/2$ comes from the properties of the Legendre $P_J$ function, which is tamest there:
\begin{eqnarray}
    P_{J}(z) &\sim_{z\to\infty}&\pi^{-1/2}\frac{\Gamma(J+1/2)}{\Gamma(J+1)}(2z)^J\quad \text{Re}\, J > -1/2\\
    &\sim_{z\to\infty}&\pi^{-1/2}\frac{\Gamma(-J-1/2)}{\Gamma(-J)}(2z)^{-J-1}\quad \text{Re}\, J \leq -1/2\,,
\end{eqnarray}
hence this choice causes the contribution arising from the background integral to be as small as possible, being the amplitude dominated by the Regge poles at large energies.


\subsection{Mandelstam-Sommerfeld-Watson transform}
\label{sec:mandelstam-regge-poles}

The previous discussion shows that the Sommerfeld-Watson transform induces a boundary in the complex $J$-plane, at $\text{Re}\, J=-1/2$, but it is not a natural boundary and Mandelstam has proposed in~\cite{MANDELSTAM1962254} a way to push the contour further away. This construction will be useful later when we want to extend our non-reggeization theorem to the non-existence of the scattering amplitude.

The basic idea is to replace the $P_J$ function with a $Q_J$ function, which has better convergence properties at infinity, and the procedure boils down to Legendre function gymnastics.\footnote{This is also what makes Froissart-Gribov powerful.}

\begin{figure}
    \centering
    \includegraphics[scale=0.9]{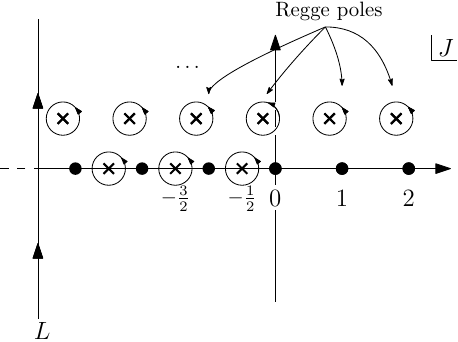} 
    \caption{Mandelstam-Sommerfeld-Watson contour. }
    \label{fig:MSW}
\end{figure}
Start from the usual partial wave expansion for the amplitude and add-and-subtract the following terms
\begin{align}
\label{eq:MandelstamSW}
A(s,t) &= 16 \pi \sum_{J=0}^\infty \left\{ (2J+1) f_J (t) P_J (z_t) 
    + \frac{1}{\pi} (2J) f_{J-1/2}(t) Q_{J-1/2}(z_t) \right\} \notag\\
&\quad - 16 \pi \sum_{J=1}^\infty \frac{1}{\pi} (2J) f_{J-1/2}(t) Q_{J-1/2}(z_t),
\end{align}
with 
\begin{equation}
\label{eq:zt-def}
    z_t=1+\frac{2s}{t-4\mu^2}\,.
\end{equation} 

As explained above in the starting point of the demonstration of the Froissart-Gribov projection, for integer $J$, $Q$ only has a cut on the $[-1;1]$ interval, with the discontinuity being the $P$ function. At non-integer $J$, $Q$ also has a cut extending from $1$ to $-\infty$.

Notice that in eq.~\eqref{eq:MandelstamSW}, the $Q$-term at $J=0$ is absent because of the factor $2J$, thus $Q_{-1/2}$ does not belong to the sum.

Now we perform a Sommerfeld-Watson transformation on the terms inside
the brackets. The first term gives the usual representation shown in
eq.~\eqref{eq:SommerfeldWatson}, while for the second one, we use
$(\cos\pi J)^{-1}$ to get the right poles at the half-integers in the
$J$-integral. Using the relation
$$\frac{P_J(z)}{\sin{\pi J}}-\frac{1}{\pi}\frac{Q_J (z)}{\cos{\pi
    J}}=-\frac{1}{\pi}\frac{Q_{-J-1}(z)}{\cos{\pi J}}, $$ we can
combine both integrals to get that the term in parenthesis in eq.~\eqref{eq:MandelstamSW} can be expressed as
\begin{equation}
   16 \pi \sum_J \bigg\{\dots\bigg\}=\frac{16}{2 i} \int_{-i \infty -1/2 + \epsilon}^{i \infty -1/2 + \epsilon}(2J+1)f_J(t)\frac{Q_{-J-1}(-z_t)}{\cos{\pi J}}\,\mathrm{d}J.
\end{equation}

We are now ready to push the contour to $\text{Re}\, J < -1/2$. We will just
pick up poles at every negative half-integer, apart from the possible
singularities of $A_J$ that we might cross. If we set the contour at
$\text{Re} \, J = - L$, we get:
\begin{multline}
    16 \pi \sum_J \bigg\{\dots\bigg\}=\frac{16}{2 i} \int_{-L-i \infty}^{-L+i \infty} (2J+1)f_J(t)\frac{Q_{-J-1}(-z_t)}{\cos{\pi J}}\,\mathrm{d}J\\-16 \pi \sum_{J=-L'}^{-1/2} \frac{(-1)^{J-1/2}}{\pi}(2J)f_{J}(t)Q_{-J-1}(-z_t) + \textrm{Regge poles and cuts},
\end{multline}
where $L'=\lceil L \rceil$ is the closest half-integer to $L$ greater than $L$. 

At half-integer points, the $Q_J$-functions possess a reflection symmetry around $J=1/2$:
\begin{equation}
    Q_J = Q_{-J-1}\,\quad J\in \mathbb{Z}+1/2\rm{.}
\end{equation}
In his original paper~\cite{MANDELSTAM1962254}, Mandelstam simply assumes that the same property holds for partial waves and that we have that $f_J = f_{-J-1}$ at half-integer $J$. This seems reasonable, from the Froissart-Gribov's perspective. Collins in his book~\cite{Collins_1977} calls this the "Mandelstam symmetry" of partial waves.  
Assuming that the Mandelstam symmetry holds, it is easy to see that by changing $J \to -J-1/2$ in the sum we will cancel the first $L'-1/2$
terms in the infinite sum outside the parenthesis in
eq.~\eqref{eq:MandelstamSW}. 

Under these conditions, we get
\begin{multline}
    A(s,t)=\frac{16}{2 i} \int_{L-i \infty}^{L+i \infty} (2J+1)f_J(t)\frac{Q_{-J-1}(-z_t)}{\cos{\pi J}}\,\mathrm{d}J\\-16 \pi \sum_{J=L'+1/2}^{\infty} \frac{(-1)^{J-1}}{\pi}(2J)f_{-J-1/2}(t)Q_{-J-1/2}(z_t) + \textrm{Regge poles and cuts}.
\end{multline}

The important point that Mandelstam makes is that for $z_t$ large, both the background
integral and the infinite sum go like $z_t^{-L}$ due to the asymptotic behaviour of the $Q_J$ (see eq.~\eqref{eq:QJasymptotic}, and thus we can suppress their contribution by pushing the contour to the left. This representation then allows us to express the amplitude exclusively in terms of the singularities of the analytical continuation of the partial waves into complex $J$ in this regime to an extremely good approximation. 

The question of whether this procedure can be continued indefinitely and the contour pushed arbitrarily to the left is an open question. This would require that the partial waves do not grow exponentially in the negative, imaginary $J$ direction. If this were the case, the contribution from the background integral and the infinite sum would vanish, and the amplitude could be exactly expressed in terms of its singularities in complex $J$ in the $t$-channel. An important point is that the terms associated with the singularities in $J$ on the $t$-channel Mandelstam-Sommerfeld-Watson (MSW) representation of the amplitude do not exhibit any of the $s$-channel singularities. Therefore, in the case in which the contour is closed at a finite value of $J$, they must be hidden in the background integral. Consequently, if we were allowed to drop this contribution this would imply a $J$-plane channel-duality, in which the $s$-channel singularities would be generated by the sum of the ones on the $t$-channel. This would provide an alternative proof to the results presented in this paper since an infinite number of Regge poles on the $t$-channel would be necessary to reproduce the $s$-channel singularities. 

A clear counter-example to this procedure is the Veneziano amplitude. As is known, and as we could observe with our previous work~\cite{Eckner:2024ggx}, the expansion over Regge poles is only asymptotic, thus the effects of the background integral must be taken into account to recover the full amplitude. However, it is still plausible to think that in some scenarios the situation could be different and that this sort of $J-$plane manipulations could be used to constrain amplitudes in a way analogous to channel duality, for instance in the present case where we have only finitely many Regge poles.

The main source of uncertainty here is that we have a priori no control over the partial waves in the background integral at large $-L'$ and it is not even clear how to define them, as the Froissart-Gribov projection is not applicable there (remember that the Mandelstam symmetry only works for \textit{integer} $J$, on the complex line above the integer it no longer holds in general).

\section{Ruling out dual model amplitudes with finitely many trajectories}
\label{sec:ruling-out}
 This section presents the main result of the paper: Meromorphic scattering amplitudes of external scalars with finitely many Regge trajectories and satisfying unsubtracted or one-subtracted dispersion relations cannot exist. Since the proof is slightly easier for the case with no subtractions, we present it first as a warm-up, even though this case is technically a sub-case of amplitudes with one subtraction (where the subtraction is trivially zero).

\subsection{Case without subtractions}

We present first the case of a dual model amplitude satisfying unsubtracted dispersion relations since the procedure is slightly more straightforward. We begin ruling out the existence of amplitudes with a single trajectory and then extend the result to any finite number.

The proof is rather simple: Channel duality imposes a very specific asymptotic behaviour for the Legendre coefficients $c_n$ at large $n$, which is incompatible with the existence of a Regge pole in the partial waves. Since a dual model amplitude comprised of a single trajectory as the one presented in eq.~\eqref{eq:dm-srt} would naturally arise in a scenario where the only singularity of the partial wave coefficients $f_J(s)$ as a function of $J \in \mathbb{C}$ is a simple pole at $J=\alpha(s)$, this strongly hints towards the inconsistency of these objects. However, we cannot exclude the possibility of an amplitude of the form of eq.~\eqref{eq:dm-srt} in which the poles in $s$ are not related by a Regge pole. Nevertheless, since we consider meromorphic amplitudes, the only allowed singularities of the partial waves are simple poles, and thus this means that the partial waves in this putative model are entire functions. This will imply in turn an incompatibility between the finiteness of the partial wave coefficients and channel duality, rendering this object inconsistent. 

\subsubsection{Decay of coefficients from channel duality}

We start by deriving the asymptotic decay condition on the $c_n$ coefficients required by channel duality. We saw in the previous section that for values of $t$ such that fixed $t$ unsubtracted dispersion are valid, relation implies that the amplitude can be expressed as a sum over $s$-channel poles as in eq.~\eqref{eq:dm-srt}. 

From the Sommerfeld-Watson transform restricted to the case in which the only singularities of the partial wave coefficients in $J$ are simple poles, we see that the large $s$ behaviour of the amplitude for any given value of $t$ must be controlled either by the leading Regge pole or by the background integral. Since the background integral depends on $s$ only through $P_J(-z)$, $z=1+2s/(t-4\mu^2)$, closing the contour at Re$J=-1/2$ as described in the previous section implies that its large $s$, fixed $t$ behaviour is $\sim s^{-1/2}$ for any $t$. Therefore, the large $s$, fixed $t$ behaviour of the full amplitude is $A(s,t) \sim \rm{max}[\,s^{-1/2},s^{\alpha(t)}\,]$. 

By definition, $\alpha(t)=0$ at $t=m_0^2$, and assuming that the trajectory functions are monotonic\footnote{This must be the case in general, as if $\alpha(t)$ was not monotonic it would either cross again the lines $J=0$ or $J=1$ at some value $t<m_0^2$, signaling the presence of a spin 0 or spin 1 state lighter than $m_0^2$ in the spectrum and thus violating the assumption that $m_0^2$ is the lightest state, or $\alpha(t)$ we would have $0<\alpha(t)<1 \, \forall t<m_0^2$, which would imply that the amplitude doesn't decay at large $s$ for any value of $t$ and therefore contradict the assumption that it satisfies unsubtracted dispersion relations. The only way out would be for the Regge residues to vanish precisely every time the trajectory function crosses integer values at $t<m_0^2$, which is a very tuned-yet not impossible- scenario.} this implies that $\alpha(t)<0, \, \forall t<m_0^2$.  Therefore, we see that $A(s,t) \to 0$ as $s \to \infty$ for $t< m_0^2$, while it goes to a constant when $t=m_0^2$, and thus unsubtracted dispersion relations hold for $t<m_0^2$.

For the $s$-channel pole expansion, this means
\begin{equation}
    \forall t<m_0^2,\quad \sum_{n=0}^{\infty}c_n \frac{P_n\!\left(1+\frac{2t}{m_n^2-4\mu^2}\right)}{\vphantom{\intop^2}m_n^2}<\infty ,
    \label{eq:convergentseriescondition}
\end{equation}
while it diverges {\it precisely} \textsc{at} $t=m_0^2$.

To analyze the condition that this imposes on the $c_n$'s, we need the asymptotic behaviour of the Legendre polynomials at large $n$, which will depend on the growth of $m_n^2$. Intuitively, near $\epsilon_n=0$, $P_n(1+\epsilon_n)$ grows like $(1+\epsilon_n)^n$, and it can either diverge if $\epsilon_n$ does not decay fast enough, go to a constant or to 1. 

It turns out that the limiting case is $n^2$. If $m_n^2$ grows slower than $n^2$ asymptotically, we get:
\begin{equation}
  \label{eq:Pn-asymptotic}
  P_n\!\left(1+\frac{2t}{m_n^2-4\mu^2}\right)\underset{n\to\infty}{\sim} \left(\frac{m_n^2}{t}\right)^{1/4} \frac{e^{2 n\sqrt{t/m_n^2}}}{2\sqrt{n \pi}}.
\end{equation}
This expression can be derived from the integral representation of the Legendre polynomials, $\frac{1}{\pi}P_n(\cos(\theta)) = \int_0^\pi (\cos(\theta)+ i
  \sin(\theta)\cos(\phi))^n\,\mathrm{d}\phi$.
The asymptotic behaviour of the $P_n(1+2t/(m_n^2-4m_0^2))$ is found by explicitly taking the limit $n \to \infty,$ $m_n^2 \to \infty$.\footnote{See also \cite{atkinson_conditions_1968}.}

For $m_n^2=n^2$, the series goes to a constant which depends on $t$:
\begin{equation}
  \label{eq:Pn-asymptotic-nsq}
  P_n\!\left(1+\frac{2t}{n^2-4\mu^2}\right)\underset{n\to\infty}{\to} {}_0{F}_1(1,t),
\end{equation}
where ${}_0{F}_1(a,z)$ is the confluent hyper-geometric function. Finally, for $m_n^2>n^2$, they go to one.

From this asymptotic behaviour, we notice two different regimes, depending on whether the spectrum is bounded asymptotically by $n^2$ or not, which produces the exponential growth, or the flat trend of order unity.

Consider first the case $m_n^2 >  n^2$ at large $n$. In this case, the Legendre polynomials simply go to 1 at large $n$ independently of the value of $t$. This means that we can effectively decompose the series in eq.~(\ref{eq:convergentseriescondition}) into finitely many terms that depend on $t$ (and are therefore regular in $t$), and an infinite $t-$independent tail of the form $\sum_n \frac{c_n}{m_n^2}$. Such a scenario can never yield a series which would be convergent for $t<m_0^2$ but diverge at $t=m_0^2$, and is thus inconsistent with crossing and channel duality.

If $m_n^2\sim n^2$ at large $n$, the same argument can be made. Since the hypergeometric function is regular, it cannot blow up at $t=m_0^2$, nor the finite sum of terms that constitute the head of the series. We reach the same conclusion.

Now, let us focus on the nontrivial case $m_n^2 < n^2$ asymptotically. For $0<t<m_0^2$ the Legendre polynomials grow exponentially as $e^{2n\sqrt{t/m_n^2}}$. Thus the only way for eq.~(\ref{eq:convergentseriescondition}) to be regular for $t<m_0^2$ is that the $c_n$ coefficients decay exponentially:
\begin{equation}
    c_n\leq \frac{e^{-2n\sqrt{m_0^2/m_n^2}}}{m_n^2}\,\rm{.}
\end{equation}

However, if the coefficients decay {\it strictly} faster than $m_0^2$, then the terms in the sum eq.~\eqref{eq:convergentseriescondition} decay exponentially when $t=m_0^2$, and therefore the series is finite at this value contradicting channel duality, which predicts that the series should diverge as $1/(t-m_0^2)$. Therefore, it must be case that the $c_n$ coefficients behave asymptotically as
\begin{equation}
\label{eq:coeff_decay_unsub}
    c_n \sim \frac{e^{-2n\sqrt{m_0^2/m_n^2}}}{m_n^2} \left(1+ \mathcal{O}(n^{-k}) \right), 
\end{equation}
for some number $k>0$.

\subsubsection{Decay of coefficients implies absence of Regge poles}

As we will see, this decay condition is too strong to accommodate the formation of a Regge pole in the partial wave coefficients. For this, we use the expression for the partial wave coefficients derived from the Froissart-Gribov projection specialized to the case of meromorphic amplitudes in eq.~\eqref{eq:FGformula-poles-residues}. In this expression, we can set $s=0$ to get rid of the Legendre polynomials via $P_n(1)=1$, to obtain
\begin{equation}
    f_{J}(0) =-\frac{i}{16\pi\mu^2} \sum_{n=0}^\infty \,Q_{J}\!\left(1-\frac{m_n^2}{2m_0^2}\right) c_n \,.
\end{equation}

If the $f_J(s)$ had a Regge pole at some $J=\alpha(s)$, then naturally $f_J (0)$ would diverge at $J=\alpha(0)$. But an even more basic question is: Can this series diverge at all? 

As mentioned before, at fixed $J$ the asymptotics of the Legendre $Q$ functions is simply $Q_J(z)\sim z^{-J-1}$, and therefore the series in the right-hand side behaves as
\begin{equation}
    \sum_{n=0}^\infty \,c_n \left(m_n^2\right)^{-J -1} \rm{.}
\end{equation}
Since the $c_n$'s are exponentially decaying and $m_n^2 < n^2$ asymptotically, the above series does not diverge for {\it any} finite value of $J$. In particular, $f_J(0)$ is perfectly regular at $J=\alpha(0)$, which proves that at least in the case without subtractions, scattering amplitudes with one trajectory cannot have a Regge pole. Since meromorphicity implies that these are the only allowed singularities in $J$ of the partial wave coefficients, we conclude that in this case \emph{partial wave coefficients are entire functions of $J$}.

As mentioned before, being this the case we lose the prediction from Regge behaviour that relates the infinitely many particles in the spectrum to each other via the function $\alpha$, and also the relation between the spectrum and the high energy behaviour. While it seems unnatural that an amplitude made up of infinitely many disconnected poles not related by any underlying structure could be compatible with crossing, in principle we cannot yet discard the possibility that an amplitude as \eqref{eq:dm-srt} could exist even in the absence of a Regge pole. However, we will see now that the absence of singularities in $J$ fn the partial waves implies the non-existence of the single-trajectory amplitude.

\subsubsection{Absence of Regge poles implies non-existence}

If the partial waves have no singularities, we can deform the contour in the Sommerfeld-Watson transform without obstructions, and the amplitude will be given entirely by the background integral:
\begin{equation}
A(s,t) = -\frac{16\pi}{2i}\int_{-1/2-i\infty}^{-1/2+i \infty}\frac{\mathrm{d}J}{\sin \pi J} (2J+1) f_J (t) P_J \left(-1-\frac{2s}{t-4 \mu^2}\right),
\end{equation}
which immediately implies that it decays at large $s$, fixed $t$ at least as fast as $s^{-\frac{1}{2}}$~\cite{MANDELSTAM1962254}. This ensures that the Froissart-Gribov projection for the partial waves can be defined for values of $J$ all the way down to $J=0$. Using this representation for the partial waves in terms of the residues derived in eq.~\eqref{eq:FGformula-poles-residues}, at least for generic $s$ not corresponding to a mass in the spectrum, the partial waves are finite and therefore
\begin{equation}
\label{eq:FGformula-convergence}
f_J(s) ={i\over4 \pi}\frac{1}{s-4\mu^2}  \sum_{n=0}^\infty c_{n}\,
    Q_J\!\left(1+\frac{2m_{n}^2}{s-4\mu^2}\right) P_{n}\!\left(1+\frac{2s}{m_{n}^2-4\mu^2}\right)\, < \infty, \quad J \geq 0.
\end{equation}

However, this is incompatible with the decay condition eq.~\eqref{eq:coeff_decay_unsub} imposed by channel duality on the $c_n$ coefficients. Choosing a value of $s$ larger than $m_0^2$ and using the asymptotic behaviour of the Legendre polynomials from eq.~\eqref{eq:Pn-asymptotic} as well as the one for the $Q$-functions and inserting the asymptotic expression for the $c_n$'s, we see that the series on the right-hand side would go at large $n$ as
\begin{equation}
\sum_n \frac{1}{\left(m_n^2\right)^{J +1}} \exp \left( 2n \frac{\sqrt{s}-\sqrt{m_0^2}}{\sqrt{m_n^2}} \right). 
\end{equation}
Since $m_n^2$ is bounded by $n^2$, the $J$-dependent contribution coming from the $Q$-function is completely negligible, and the elements in the sum blow up exponentially as $c_n \sim e^{2(\sqrt{s}-m_0)}$. The only way in which the partial wave coefficients would be well defined would be if the $c_n$'s satisfied
\begin{equation}
\label{eq:coeff_decay_condition_FG}
\lim_{n \to \infty}
c_n \, e^{2n \sqrt{\frac{s}{m_n^2}}}=0,
\end{equation}
for positive $s$ as large as we want. This is clearly in contradiction with the conditions imposed by crossing symmetry and channel duality on the $c_n$ coefficients, which leads us to conclude that dual model amplitudes satisfying unsubtracted dispersion relations and composed of a single trajectory as in \eqref{eq:dm-srt} are inconsistent with the bootstrap principles and therefore \emph{are ruled out}.

\subsubsection{Generalization to finitely-many trajectories}
\label{sec:finite-RP}

The generalization of the previous reasoning to finitely many trajectories is straightforward. An amplitude consisting of $N+1$ Regge trajectories can be written as
\begin{equation}
    \label{eq:multipletrajamplitude}
    A(s,t) = \sum_{i=0}^{N} \sum_{n=0}^\infty c_{n,n-i}\frac{P_{n-i}\!\left(1+\frac{\vphantom{\intop^2}2t}{\vphantom{\intop^2}m_{n,n-i}^2-4\mu^2}\right)}{\vphantom{\intop^2}s-m_{n,n-i}^2},
\end{equation}
where $i$ indexes the different trajectories starting from the leading one at $i=0$, as in fig.~\ref{fig:dual-model}.

Similarly, the equation for the partial wave coefficients in the case of $N+1$ trajectories is given by
\begin{equation}
\label{eq:FGformula-multipletraj}
    f_J(s) ={i\over4 \pi}\frac{1}{s-4\mu^2} \sum_{i=0}^{N}\sum_{n=0}^\infty \,Q_J\!\left(1+\tfrac{2m_{n,n-i}^2}{\vphantom{\intop^2}s-4\mu^2}\right) c_{n,n-i} P_n\!\left(1+\tfrac{\vphantom{\intop^2}2s}{\vphantom{\intop^2}m_{n,n-i}^2-4\mu^2}\right)\,.
\end{equation}

It is important to note that although the partial wave coefficients should have poles at the values $\left.J=\alpha_i(s)\right|_{i=0,\dots,N}$ corresponding to the different trajectories, the expression given in eq.~(\ref{eq:FGformula-multipletraj}) only makes sense until we hit the first singularity in $J$, and therefore we can only use it to probe the leading Regge pole at $J=\alpha_0(s)$, where the series is expected diverge. Beyond this value of $J$ we are outside of the region of convergence of the series, and therefore we should not expect that the partial wave coefficients are correctly described by it.

The implications of channel duality are very similar to the previous, single trajectory case. The series should still converge up to $t=m_{0,0}^2=:m_0^2$. As before, we need to distinguish the cases $m_{n,n}^2 \geq n^2$ and $m_{n,n}^2 < n^2$ when $n \to \infty$. The generalization for many trajectories with spectra growing faster than quadratically is trivial, as they fail for the same reason as in the single trajectory and the argument does not require any modifications.

For $m_n^2$ spectra growing slower than $n^2$, the argument generalizes easily. Along each trajectory, the coefficients must still decay to compensate for the exponential growth of the Legendre polynomials and ensure that 
\begin{equation}
    \sum_{n=0}^{\infty}c_{n,n-i} \frac{P_{n-i}\!\left(1+\frac{2t}{\vphantom{\intop^2}m_{n,n-i}^2-4m_0^2}\right)}{\vphantom{\intop^2}m_{n,n-i}^2}<\infty, \quad t<m_0^2, \quad i=0,\dots,N\rm{,}
    \label{eq:convergentseriesconditionmanytraj}
\end{equation}
so that we get
\begin{equation}
    c_{n,n-i}\leq e^{-2(n-i)\sqrt{m_0^2/m_{n,n-i}^2}}\rm{.}
\end{equation}

The main difference with the case of a single Regge trajectory is that now we don't get a strict decay condition as~\eqref{eq:coeff_decay_unsub} on all the trajectories, but we find that an analogue decay must be satisfied by at least one of them to generate the pole at $t=m_0^2$. To see this one simply needs to note that as $t \to m_0^2$, the sum of the contributions from each trajectory as the one in eq.~\eqref{eq:convergentseriesconditionmanytraj} must diverge as $(t-m_0^2)^{-1}$. However, since the sum over trajectories has a finite number of terms, the full sum can only diverge if at least one of its constituents diverges, although not necessarily all of them must do so. Therefore, the decay condition on the coefficients in the case of many trajectories is
\begin{equation}
    c_{n,n-i}\leq e^{-2(n-i)\sqrt{m_0^2/m_{n,n-i}^2}}, \quad \forall i=0,\dots,N, 
\end{equation}
and
\begin{equation}
    c_{n,n-k}\sim e^{-2(n-k)\sqrt{m_0^2/m_{n,n-k}^2}}, \quad \text{for at least one }k \in 0,\dots,N.
\end{equation}

A similar argument can be used to show that, as in the single-trajectory case, these decay conditions are too strong to allow the existence of Regge poles.

Imagine that there existed a Regge pole associated with the leading trajectory. Then at least for $\text{Re}(J)>\alpha_0(s)$ the series is convergent, and thus in this region we are free to re-organize the terms in the sum as we want, and we can express the partial wave coefficient as
\begin{equation}
\label{eq:PWcoef-multipletraj}
    f_J(s) =\sum_{i=0}^{N}f_J^{(i)}(s),
\end{equation}
with 
\begin{equation}
\label{eq:PWcoefcomponents}
    f_J^{(i)}(s) ={i\over4 \pi}\frac{1}{s-4\mu^2} \sum_{n=0}^\infty \,Q_J\!\left(1+\tfrac{2m_{n,n-i}^2}{s-4\mu^2}\right) c_{n,n-i} P_{n-i}\!\left(1+\tfrac{2s}{m_{n,n-i}^2-4\mu^2}\right)\,,
\end{equation}
being the contribution to the partial wave coefficient provided by the $i^{\text{th}}$ trajectory.
Again, given that the sum over trajectories has a finite number of terms, demanding that $f_J(s)$ diverges as $J\to \alpha_0 (s)$ implies that at least one of the components $f_J^{(i)}(s)$ must diverge at this value of $J$. 
Repeating the argument used for the single-trajectory case, we see that at $s=0$ the exponential damping of the $c_{n,n-i}$ coefficients yields partial wave trajectory coefficients $f_J^{(i)}$ that are regular and therefore the partial wave coefficients are free of poles also in this case.

The final step showing that this implies an inconsistency in the partial wave coefficients goes in the same way as in the single trajectory case. The absence of Regge poles in the case of finitely many trajectories again implies that the Froissart-Gribov representation holds for $J\geq 0$, and in turn, this implies that to have non-divergent partial wave coefficients the Legendre coefficients must satisfy a property analogous to eq.~\eqref{eq:coeff_decay_condition_FG}, namely
\begin{equation}
\label{eq:coeff_decay_condition_FG_multitraj}
\lim_{n \to \infty}
c_{n,n-i} \, e^{2(n-i) \sqrt{\frac{s}{m_{n,i}^2}}}=0, \quad \forall i=0,\dots,N,
\end{equation}
again for positive $s$ as large as desired. Since this is in contradiction with the decay condition required to have channel duality, we conclude that the no-go also holds in the case of finitely many trajectories, and therefore these objects are ruled out too.

A crucial point is that the previous argument fails if we had an infinite number of Regge trajectories: in this case, it would be possible that the contribution from each trajectory is finite, but the infinite sum over trajectories could give rise to the divergence of the partial wave coefficients. We will show later that this is precisely what happens for the Veneziano amplitude. 

Importantly, this means that the residues corresponding to the subleading trajectories could in principle be arbitrarily small: this would not prevent the infinite sum from yielding a pole in $J$ in $f_J(s)$. Therefore, the argument we presented here does not put any constraint on the absolute size of the couplings $c_{n,J}$, and thus there is no immediate clash with the numerical results of~\cite{Albert:2023seb,Albert:2022oes,Albert:2023jtd}. The seemingly single-trajectory extremal amplitude that they observe could very well be a limiting case in which the couplings of the subleading trajectories are much smaller than those of the leading one and therefore are not visible due to finite numerical precision.

\subsubsection{Extension to higher dimensions}
\label{sec:higher-d}

The extension of the previous result to higher dimensions can be proved as follows. In higher dimensions $D=4+d$, Legendre polynomials get replaced by Gegenbauer polynomials with extremely similar properties. In order to rule out those cases as well, it is sufficient to observe that if meromorphic amplitudes with finitely many trajectories existed, they could be dimensionally reduced to $D=4$ dimensions and would give rise to amplitudes with a larger number of Regge trajectories but still finitely many. Since these are excluded, they cannot have an ancestor in $4+d$ dimensions and therefore these amplitudes cannot exist either, extending the previous results to $D\geq4$.

\subsubsection{How does the Veneziano amplitude evade the no-go?}
\label{sec:Veneziano}
As mentioned before, in the case of infinitely many Regge trajectories the exponential decay of the $c_{n,J}$ coefficients does not necessarily clash with the existence of Regge poles, as the divergence of the partial wave coefficients can be caused by the sum of the (finite) contributions of the infinitely many trajectories and not by the divergence of the contribution from a single trajectory. This is good news, as we would be in trouble if we found a constraint inconsistent with the existence of the Veneziano amplitude.

To check that this is the case, we use the fact that the residues of the Veneziano amplitude are given by
\begin{equation}
    \label{eq:ResVeneziano}
    R_n^{V}(s)=\frac{(-1)^n}{n!} \frac{\Gamma(m_0^2-s)}{\Gamma(m_0^2-s-n)}.
\end{equation}

Using Stirling's approximation for the Gamma function and the factorial as well as the Euler reflection formula of the Gamma function, it is simple to show that their asymptotic behaviour at large $n$ is given by
\begin{equation}
    \label{eq:ResVenezianoAsymptotic}
    R_n^{V}(s)\sim n^{s-m_0^2}, \quad n \to \infty.
\end{equation}

Plugging this result for the residues and the asymptotic behaviour of the $Q_J$ functions into eq.~(\ref{eq:FGformula-poles-residues}) we see that
\begin{equation}
    {f_J^{V}(s)}\sim \sum_n n^{-J-1+s-m_0^2},
\end{equation}
which diverges as a simple pole precisely at the leading Regge pole $J=s-m_0^2$. To achieve this we used the full residue at each pole in $s$, and therefore all the infinitely many trajectories contributed to generating the divergence of the partial wave coefficient, as can be seen from decomposing the residues in Legendre polynomials. 
Note that the coefficients in the Legendre decomposition of the residues of the Veneziano amplitude are exponentially decaying at large $n$, meaning that the contributions of the form of eq.~(\ref{eq:PWcoefcomponents}) are convergent when evaluated in the case of a linear spectrum. It is the infinite sum over trajectories that causes $f_J$ to diverge.\footnote{See some related comments in~\cite{Nayak:2017qru}.}

\subsection{Case with one subtraction}

We will now repeat the proof in the case of amplitudes satisfying once-subtracted dispersion relations. For this case, the most important differences are the modification of the channel duality condition due to the presence of the subtraction term and the fact that we cannot set $s=0$ in the partial waves anymore to remove the Legendre polynomials from the Froissart-Gribov formula, since in this case there might exist a fixed pole at $J=0$ arising from a constant term in the amplitude and thus we cannot assume the series expansion for the partial waves to converge at $J=\alpha(0)<0$. It will however suffice to set it to a value $s_0\in\left(0, m_0^2\right)$, for the regularity of the partial waves to be implied by the regularity of the amplitude in $t$.

\subsubsection{Modified channel duality}

As was mentioned in the previous section, for amplitudes requiring one subtraction dispersion relations do not allow for expressing the amplitude exclusively as a sum over $s$-channel poles as in the unsubtracted case, but rather the expression~\eqref{eq:DR-1sub} involving explicit singularities in both channels.
The notion of channel duality is modified in the light of this expression, as now the leading $t$-channel pole at $t=m_0^2$ appears explicitly in the sum, and therefore we do not need to require the infinite sum in eq.~\eqref{eq:DR-1sub} to diverge at this value of $t$. In fact, by subtracting from it the pole at $t=m_0^2$, we can extend the region for which the series is convergent to $t< m_1^2$. Specializing to the case of a single trajectory, i.e.~with no sum on $J$ but simply $J=n$ and our already established notation $m_{n}:=m_{n,n},c_n:=c_{n,n}$, subtracting the leading $t$-channel pole yields
\begin{equation}
\label{eq:subtracted_duality}
    A(s,t) -  \frac{c_0}{t-m_0^2}-c_0
    =
    \sum_{n,j} \frac{c_{n}}{m_{n}^2}\left( \frac{s\, P_n\!\left(1+\tfrac{\vphantom{\intop^2}2t}{\vphantom{\intop^2}m_{n}^2-4\mu^2}\right)}{\vphantom{\intop}s-{m_{n}^2}} + \frac{t}{t-{m_{n}^2}}(1-\delta_{0,n}) \right)\rm{.}
\end{equation}
In this expression, it is clear that the left-hand side has to be regular at $t=m_0^2$, and thus it must be so for the right-hand side. As we already noted, the singularity at $t=m_1^2$ is reproduced by the combination of the explicit pole in $t$ and the infinite sum in $s$ to produce the correctly normalized Legendre polynomial $P_1 \left(1+\frac{2s}{m_1^2-4 \mu^2} \right)$ for the residue at $t=m_1^2$.

\subsubsection{Case with a single trajectory}

We can now repeat arguments very similar to those of the un-subtracted case. Firstly, for spectra growing faster or as fast as $n^2$, $m_n^2\geq n^2$, just as before, the $t$-dependence of the term 
$$\frac{s\, P_n\!\left(1+\tfrac{\vphantom{\intop^2}2t}{\vphantom{\intop^2}m_{n}^2-4\mu^2}\right)}{s-{m_{n}^2}},$$ 
drops out at large $n$ and therefore it cannot produce a singularity in $t$ at all, hence the expected pole with correctly normalized, and $s$-dependent residue
$$c_1\frac{P_1\!\left(1+\frac{\vphantom{\intop^2}2s}{\vphantom{\intop^2}m_1^2-4\mu^2}\right)}{t-m_1^2},$$ cannot be generated. Thus, these spectra are excluded by crossing and subtracted channel duality alone.

Let us now turn to spectra growing slower than $n^2$ asymptotically, $m_n^2<n^2$. For them, demanding that after removing the pole at $t=m_0^2$ the once-subtracted $s$-channel expansion of the amplitude should converge up to $t=m_1^2$ and diverge precisely at $t=m_1^2$ also forces an exponential decay of the coefficients $c_n$ by the same reasoning as in the previous section. The asymptotic behaviour of the $c_n$ at large $n$ in this case is of the form
\begin{equation}
    \label{eq:decay-1sub}
    c_n \sim \frac{e^{-2n \sqrt{m_1^2/m_n^2}}}{m_n^2}\,.
\end{equation}

We will now see that again this asymptotic decay prevents a Regge pole from being generated in the partial wave coefficients via the Froissart-Gribov formula. The discussion differs from the one in the unsubtracted case, since in this case 
we need to deal with a subtle point regarding the analyticity in $J$ for the once-subtracted amplitude. 
The need for one subtraction implies the presence of a constant term in the amplitude, which would generate the divergence of the integral in the Froissart-Gribov projection at $J=0$ even for negative values of $t$. This implies that the subtraction yields a pole at $J=0$ in the complex $J$-plane, sometimes referred to as a \textit{fixed} Regge pole.

If the $f_J$ has singularities, the representation given in eq.\eqref{eq:FGformula-poles-residues} would break down as soon as $J$ reaches the first one, and therefore we must restrict to values of $J$ to the right of this putative pole in the complex $J$-plane when using it.
Since $\alpha(m_n^2)=n$,\footnote{Remember that this follows by the very definition of our leading and single trajectory, a state of mass $m_n^2$ has spin $n$.} we have $\alpha(m_0^2)=0$ and therefore we need to analyze the behaviour of the partial wave $f_J(s)$ at some value of $s>m_0^2$, where the putative Regge pole and fixed pole would coincide at $J=0$. In particular, the previous choice $s=0$ would not work. 

Let us then choose $s=s_0$, with $m_0^2<s_0<m_1^2$ and look at $f_J(s_0)$, given by:

\begin{equation}
    f_J(s_0) ={i\over4 \pi}\frac{1}{s_0-4\mu^2} 
    \sum_{n=0}^\infty c_{n} \,Q_J\!\left(1+\tfrac{\vphantom{\intop^2}2m_{n}^2}{\vphantom{\intop^2}s_0-4\mu^2}\right) P_n\!\left(1+\tfrac{\vphantom{\intop^2}2s_0}{\vphantom{\intop^2}m_{n}^2-4\mu^2}\right)\,.
\end{equation}
In this formula, the $c_n$'s decay exponentially as eq.~\eqref{eq:decay-1sub} and the $Q_J$ behaves as a power of $m_n^2$. Since $s_0<m_1^2$, the exponential growth of the $P_n$'s given by 
\begin{equation}
    P_n\!\left(1+\tfrac{\vphantom{\intop^2}2s_0}{\vphantom{\intop^2}m_{n}^2-4m\mu^2}\right)\sim \frac{e^{2n \sqrt{s_0/m_n^2}}}{m_n^2},
\end{equation}
is not strong enough to counterbalance the decay of the $c_n$'s which still yields an overall exponentially decaying behaviour, and we conclude just as before that the partial waves don't have poles in $J$.

The rest of the story is identical to the case without subtractions, with the only difference that the fixed pole at $J=0$ means that we can only require the convergence of the Froissart-Gribov projection for $J\geq1$ instead as $J\geq0$ as was the case for amplitudes satisfying unsubtracted dispersion relations. However, the role of $J$ in the analysis of eq.~\eqref{eq:FGformula-convergence} in the unsubtracted case was irrelevant as it only enters the expression for the $f_(s)$ as a power, with the convergence or divergence of the sum being determined by the relative weights of the $c_n$ coefficients and the Legendre polynomial exponential behaviours. Therefore, we conclude that \emph{single trajectory amplitudes are also ruled out in the case with one subtractions}.

The extension to finitely many trajectories and higher dimensions just follows through in exactly the same way as for the un-subtracted case, simply modifying the channel duality condition to consider the subtraction and dealing with the fixed Regge pole as was just shown.

An interesting remark is that objects of the form $A(s,t) \sim \frac{1}{(s-m^2)(t-m^2)}$ are known to yield consistent, unitary amplitudes at tree-level, and thus should not be ruled out by the previous arguments. Indeed they are not, and while amplitudes of this form satisfy unsubtracted dispersion relations (and naturally also the subtracted one~\eqref{eq:DR-1sub}) they evade this no-go theorem because the arguments presented before break down when all the states lie at exactly the same mass. This is not surprising, as in this scenario even the notion of single-trajectory stops making sense, with the Regge trajectory becoming vertical and thus not corresponding to a well-defined function.

While rather obvious, it is worth commenting that other simple tree-level amplitudes satisfying once-subtracted dispersion relations that evade our result are amplitudes of the form $A(s,t) \sim \frac{1}{s-m^2}+\frac{1}{t-m^2}$. This case is even simpler, as there is only one spin 0 state exchanged and therefore the framework of Regge trajectories does not apply. The upshot is that amplitudes with finitely many trajectories and a spectrum containing infinitely many different masses, as would be the case in the presence of a Regge pole, are inconsistent, while the simple tree-level amplitudes with finitely many poles that would arise in QFT are of course not.



%
%

\section{Dual bootstrap for amplitudes with one subtraction}
\label{sec:linear-programming}

To complement the results from the previous section, here we present a numerical method to study amplitudes with one Regge trajectory directly in momentum space.
In particular, we want to relate to the works of~\cite{Albert:2022oes,Albert:2023jtd,Albert:2023seb}. To this end, we introduce a simple dual numerical bootstrap inspired by the seminal work~\cite{Rattazzi:2008pe}.  It uses linear programming to rule out that such amplitudes could have only one trajectory, both in the un- and once-subtracted case. While by the above arguments this possibility was excluded already, it is our belief that the numerical approach not only serves as a cross-check, but could also be extended to study general dual model amplitudes without assuming a given number of trajectories, and therefore we chose to include it to make the method available. 

\subsection{Setup and analysis framework}
\label{sec:dual-bootstrap-setup}
We start from the fixed-$t$ once-subtracted dispersion relation written in~\eqref{eq:DR-1sub}, which we reproduce here for convenience
\begin{equation}
    A(s,t)=A(0,0)+\sum_{n=0}^{\infty}c_n\left(\frac{s \, P_n\!\left(1+\frac{2t}{\vphantom{\intop^2}m_n^2-4\mu^2}\right)}{m_n^2(s-m_n^2)} +\frac{t}{m_n^2(t-m_n^2)} \right),
\end{equation}
where $\mu^2$ is the mass squared of the external particle, $A(0,0)$ is an undetermined constant, corresponding to the value of the amplitude at $t=s=0$.  As discussed in the previous section around eq.~\eqref{eq:subtracted_duality}, upon subtracting the leading singularity in $t$ the infinite series converges for $t$ up to $t=m_1^2$.

Naturally, a fixed $s$ dispersion relation on the $t$-channel would yield an analogue expression that converges for $s<m_1^2$, and thus in the region $s<m_1^2$, $t<m_1^2$ we can write the crossing equation in terms of the following blocks
\begin{equation}
  \label{eq:crossing-blocks}
  F_n(s,t):=\frac{s\left(P_n\!\left(1+\frac{2t}{\vphantom{\intop^2}m_n^2-4m_0^2}\right)-1\right)}{\vphantom{\intop^2}m_n^2(s-m_n^2)} - \frac{t\left(P_n\!\left(1+\frac{2s}{\vphantom{\intop^2}m_n^2-4m_0^2}\right)-1\right)}{\vphantom{\intop^2}m_n^2(t-m_n^2)}
\end{equation}
as
\begin{equation}
\label{eq:crossing-Fn}
    \sum_{n=0}^\infty c_n F_n(s,t)\equiv0\rm{,}
\end{equation}
where we repeat that the $c_n$ coefficients must be non-negative as a consequence of unitarity. Therefore, eq.~\eqref{eq:crossing-Fn} is an infinite sum over the functions $F_n(s,t)$ with positive coefficients $c_n$, and one can attempt to show the absence of solutions by acting on it with linear functionals $\Lambda$.

Let us recall the geometric interpretation given in \cite{Rattazzi:2008pe} to illustrate the procedure. The quantities $F_n$ can be seen as elements of an abstract vector space of functions of the Mandelstam variables $s$ and $t$ determined by the choice of spectrum $\{m_n^2\}_{n=0}^{\infty}$. Hence, eq.~\eqref{eq:crossing-Fn} can be read as stating that a linear combination of vectors multiplied by positive positive coefficients $\{c_n\}_{n=0}^{\infty}$ sums to zero. Such a linear combination of vectors may or may not exist within the vector space. If we can construct a functional satisfying the condition that
\begin{equation}
  \label{eq:exclusion-hypothesis}
  \Lambda(F_n) \geq 0\quad\forall\;n,
\end{equation}
it is straightforward to see that crossing symmetry, eq.~\eqref{eq:crossing-Fn} can never hold for the specific set of $F_{n}$ corresponding to this spectrum,\footnote{Of course, it is necessary that at least one $\Lambda(F_n)$ is strictly positive to avoid the trivial solution.} and thus a single-trajectory dual amplitude with this set of masses is incompatible with crossing symmetry and unitarity, independently of the values of the coefficients $c_n$. 

Using this approach, if one could parameterize and scan the space of all possible spectra, and search for linear functionals that rule them out, one could settle the question of the non-existence of amplitudes with finitely many trajectories directly without invoking Regge theory. 
While we have not performed this task fully, we have observed that for all the possible spectra that we tried it was always possible to find a functional that excludes them, hinting at the inconsistency of these amplitudes with crossing symmetry and unitarity.

In the following subsection, we present a functional which excludes an amplitude which would be made of only the leading trajectory of~\cite{Albert:2023seb}.

\subsection{Impossibility of having only one trajectory in pion scattering at large $N$}

We briefly recall the context we already mentioned above. In a series of papers, the authors of \cite{Albert:2022oes,Albert:2023jtd,Albert:2023seb} used modern dispersive numerical bootstrap techniques to constrain the scattering of pions in large-$N$ theories and found an extremal solution which appears to contain a full Regge trajectory whose light states have masses remarkably close to those of real-world QCD. While they also find other resonances apart from the ones in the leading Regge trajectory, the couplings between the pion and the particles on the trajectory are orders of magnitude larger than the couplings with particles that do not lie on it. This prompted speculation about whether this extremal solution could correspond to an amplitude characterized by a single Regge trajectory, thereby suggesting that the additional apparent states might be spurious.

We will now explain how the dual bootstrap presented in sec.~\ref{sec:dual-bootstrap-setup} can rule out this possibility. 
Let us present our functional. We work with a class of functionals given by differential operators of the form 
\begin{equation}
    \label{eq:differentialoperator}
    \Lambda_{\text{diff}}\left(F_{n}(s,t)\right)=\sum_{a=2}^{N_{\text{max}}}\sum_{b=1}^{a-1}\lambda_{ab}\frac{\partial^{a+b}}{\partial s^{a}\partial t^b}F_n (s,t)\bigg|_{s=t=0},
\end{equation}
where the $\lambda_{ab}$ are constant coefficients that specify the functional, and $N_{\text{max}}$ is the maximum number of derivatives taken with respect to $s$. For these functionals, the positivity condition eq.~\eqref{eq:exclusion-hypothesis} becomes an infinite set of linear constraints on the parameters $\lambda_{ab}$, and thus the question of finding a suitable functional reduces to a linear programming problem in which one scans over the space of $\lambda_{ab}$ parameters for points inside the region circumscribed by these constraints.

An advantage of these functionals is that for a given spectrum we can obtain analytical expressions for any number of derivatives of the $F_n$ at $s=t=0$:
\begin{equation}
    \label{eq:Fnderivab}
   F_n^{ab}:= \frac{\partial^{a+b}}{\partial s^{a}\partial t^b}F_n (s,t)\bigg|_{s=t=0}={\frac{(1-a+n)_{2a}}{(m_n^2)^{b+1}(m_n^2-4\mu^2)^a}\frac{b!}{a!}-\frac{(1-b+n)_{2b}}{(m_n^2)^{a+1}(m_n^2-4\mu^2)^b}\frac{a!}{b!}}\rm{,}
\end{equation}
where $(x)_{n}$ denotes here the Pochhammer symbol $(x)_n:=x(x+1)\dots(x+n-1)$.

For a power law spectrum $m_n^2 \sim n^{1+\gamma}$ we can straightforwardly see from eq.~\eqref{eq:Fnderivab} that the asymptotic behaviour of the $F_n^{ab}$ at large $n$ is given by
\begin{equation}
    \label{eq:Fnderivab_asymptotic}
    F_n^{ab} \sim \frac{b!}{a!}n^{2a-(1+\gamma)(a+b+1)}, \quad n \to \infty,
\end{equation}
and therefore for $-1<\gamma<1$, meaning spectra bounded between constants and $n^2$:
\begin{equation}
    {\rm cst}\leq m_n^2 \leq n^2\,,
\end{equation}
and for large enough $n$, the term $F_n^{N_{\text{max}},0}$ will dominate over all the other contributions to the action of the functional on $F_n(s,t)$. Therefore, setting $\lambda_{N_\text{max},0}>0$ suffices to guarantee the positivity of the functional at large $n$. This allows us to impose the positivity of the functional only up to a finite value of $N_{\rm max}$, as long as it is large enough for the $F_n^{ab}$ to enter the regime of their asymptotic behaviour, rendering the problem finite-dimensional.

\begin{figure}
    \centering
    \includegraphics[scale=0.8]{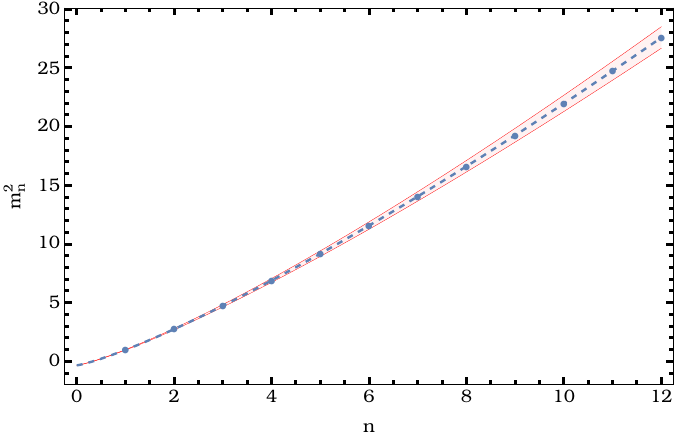}
    \caption{Blue circles: Spectrum extracted from \cite{Albert:2023seb}, Dashed blue line: Fitted function $m_n^2=-0.33+1.32 \, n^{1.23}$, Red band: Region of spectra ruled out by the functional.}
    \label{fig:pion-spectrum}
\end{figure}

The spins and masses (in units of the $\rho$ meson mass) of the particles in the putative single trajectory are found in the table on page 20 of \cite{Albert:2023seb}. These values can be fitted extremely well by a power-law spectrum given by $m_n^2=a+b \, n^{c}$, with $a=-0.33$, $b=1.32$ and $c=1.23$, as is shown in fig.~\ref{fig:pion-spectrum}.

To account for possible fitting errors, and in order to exclude a continuous set of parameters rather than a unique amplitude, we defined a band around the fitted spectrum by allowing for a tolerance of $\pm 0.01$ in the parameters $a$, $b$, $c$ (see fig. \ref{fig:pion-spectrum}), and we searched for a functional that rules out all spectra residing inside the band. 
We found a functional with $N_{\text{max}}=6$ derivatives that excludes spectra of the form $m_n^2=a+b \, n^{c}$ with $a=a_{\text{min}}+(a_\text{max}-a_\text{min})\lambda$, $b=b_{\text{min}}+(b_\text{max}-b_\text{min})\lambda$, $c=c_{\text{min}}+(c_\text{max}-c_\text{min})\lambda$, with $\lambda \in [0,1]$ and $a_{\text{max/min}}=-0.33 \pm 0.01$, $b_{\text{max/min}}=-1.32 \pm 0.01$, and $c_{\text{max/min}}=1.23 \pm 0.01$. 
To this end, we uncovered a solution to the set of
linear constraints on the $\lambda_{ab}$ generated by requiring positivity of the functional up to $n_\text{max} = 500$
for a grid of ten values of $\lambda$ in the range $[0,1]$ and the asymptotic positivity condition $\lambda_{6,0}>0$. 

\begin{figure}
    \centering
    \includegraphics[scale=1]{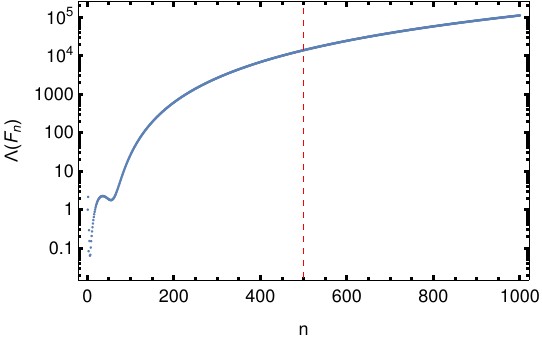}
    \caption{Action of the functional on the spectrum fitted from \cite{Albert:2023seb} as a function of $n$. The red dashed line marks the value $n_{\text{max}}=500$ up to which positivity was imposed on the functional.}

    \label{fig:pion_functional}
\end{figure}

\textit{A posteriori}, we verified that the functional does not only exclude the enforced values of $\lambda$ but also those with any value on $\lambda$ inside the desired range, and thus any spectrum that is contained inside the error band shown in fig.~\ref{fig:pion-spectrum}. As an illustrative example, the action of the functional as a function of the level $n$ in the case of the fitted spectrum is shown in fig.~\ref{fig:pion_functional}.

The coefficients of the functional are given in Table~\ref{tab:coefs}.
\begin{table}[]
\centering
\begin{tabular}{|l|l|l|l|l|l|l|}
\hline
$\lambda_{2,1}$ & $\lambda_{3,1}$ & $\lambda_{3,2}$ & $\lambda_{4,1}$ & $\lambda_{4,2}$ & $\lambda_{4,3}$ & $\lambda_{5,1}$ \\ \hline
-159.823        & 462.148         & 3933.43         & -446.939        & -19708.7        & -30857.1        & 143.114        \\ \hline
\end{tabular}

\vspace{0.5cm}

\begin{tabular}{|l|l|l|l|l|l|l|l|}
\hline
$\lambda_{5,2}$ & $\lambda_{5,3}$ & $\lambda_{5,4}$ & $\lambda_{6,1}$ & $\lambda_{6,2}$ & $\lambda_{6,3}$ & $\lambda_{6,4}$ & $\lambda_{6,5}$ \\ \hline
13988.6         & 352277          & 280339          & 1               & 19092.9         & -638357         & 769317          & -208664         \\ \hline
\end{tabular}
\caption{Coefficients of the functional that excludes amplitudes within the red sheaf of fig.~\ref{fig:pion-spectrum}.\label{tab:coefs}}
\end{table}

\section{Discussion}
\label{sec:discussion}

In this paper, we have proven solely using the analytic structure of the scattering amplitude that scalar meromorphic amplitudes with finitely many Regge trajectories, with zero and one subtraction are not consistent with analyticity, unitarity and crossing symmetry in the form of channel-duality, and therefore cannot exist. 

As noted in the introduction, we realized after completion of the proof presented here that the question of the compatibility of reggeization with a finite number of trajectories had been investigated in the literature before~\cite{mandula-slansky-singletraj, Ademollo:1968cno, atkinson_conditions_1968, Fujisaki:1970ce}. These studies all rely on the use of the construction known as FESRs to prove the non-existence of single-trajectory objects (we review FESRs in sec.~\ref{app:fesr}). In addition, they do not work out the case with subtractions, crucial to make contact with~\cite{Albert:2022oes,Albert:2023jtd,Albert:2023seb}. Lastly, the use of FESRs presents several hidden assumptions, for which no justification was given in the literature to the best of our knowledge. Let us discuss these points in more detail.

\subsection{Relation to previous works}
\label{sec:history}

\paragraph{Use of FESRs}

The common assumption of these early works is that the scattering amplitude under scrutiny exhibits Regge behaviour. The authors of \cite{mandula-slansky-singletraj} were the first to bring forward an argument. Their reasoning is rather sketchy and invokes a manipulation of the relevant FESRs that is hard to justify. In fact, the authors of \cite{atkinson_conditions_1968} (and independently \cite{PhysRevLett.21.383}) criticized this first paper deeming some of the made claims to be wrong. Instead, they develop their own but similar line of reasoning that appears more mathematically sound and leads to the exclusion of single-trajectory amplitudes with polynomially bounded spectra (the case of linear spectra was independently considered and excluded in \cite{Ademollo:1968cno}).  The crucial point is the difference between FESRs evaluated at a certain level $N$ and the subsequent level $N+1$. This difference should result in a separable expression in spin $J$ of the resonance and level $N$ in order to be compatible with the Regge behaviour which is not obtained. The third work \cite{Fujisaki:1970ce} was not aware of \cite{atkinson_conditions_1968} but references \cite{mandula-slansky-singletraj}. The latter work lacks a clear statement about the possibility of finitely many Regge trajectories, which the author of \cite{Fujisaki:1970ce} addresses. The proof is similar to \cite{atkinson_conditions_1968} in style but the final compatibility condition relies on derivatives rather than a finite difference. This work rules out reggeizing scattering amplitudes with any finite number of trajectories. Yet, it is not fully evident how rigorous these proofs really are since they rely on the FESRs, which could only hold for two limiting processes:\emph{(i)} the high-energy limit $s\rightarrow\infty$ and \emph{(ii)} and $R\rightarrow\infty$ levels considered in the contour integral. 
Finally, these works are restricted to the case with exact duality (no subtraction).

Our work therefore supplements the existing body of literature with an argument directly tied to the amplitude's properties, with no extra assumptions. In addition, we extended the proof to the subtracted case, which is the relevant one for pion scattering in large N QCD and therefore to make contact with~\cite{Albert:2022oes, Albert:2023jtd, Albert:2023seb}.

\paragraph{Claims about spurious poles for non-equal mass scattering.}

Another set of works in the literature~\cite{Freedman:1966zec, Nachtmann:2003ik}~\footnote{More precisely, see section 2.7 of~\cite{Nachtmann:2003ik} for a modern account of the argument.} claims to have identified unphysical, spurious poles occurring at $t=0$ in the contribution from a Regge pole to the Sommerfeld-Watson transform in the case of scattering of particles with unequal masses and postulates the need for infinitely many so-called {\it daughter} trajectories to cancel their contributions.  We briefly mention the origin of these would-be poles here, but remark that the proof is inconclusive as it relies on the possibility of removing the background integral by shifting it to the left indefinitely, which we argue in this paper is in general not possible. In this case, the spurious poles could simply be cancelled by the background integral contribution, rendering the argument not valid in general.

Let us explain briefly the reasoning.
When the scattering involves particles of different masses, the expression for $z=\cos \theta$ becomes more complicated, and in the $t$-channel, it reads
\begin{equation}
    z_t = 1 + \frac{s t}{T(t)}, \quad 2T(t)=t^2 - 2t(m_1^2+m_2^2)+(m_1^2-m_2^2)^2,
\end{equation}
in the case where two different masses $m_1$ and $m_2$ are involved. When this expression is substituted in the contribution from a Regge pole $\alpha_0 (t)$ to the Sommerfeld-Watson transform, the high energy expansion of the Legendre polynomial reads
\begin{equation}
\label{eq:noneqmasses}
   P_{\alpha_0(t)}\left(-1-\frac{st}{T(t)}\right)\sim \left(\frac{t}{T(t)}\right)^{\alpha_0(t)}s^{\alpha_0 (t)}\left(c_0  + c_1 \frac{T(t)}{t}\frac{1}{s}+c_2 \left(\frac{T(t)}{t}\right)^2 \frac{1}{s^2}+ \dots\right),
\end{equation}
which has singularities at $t=0$ and $t=(m_1 \pm m_2)^2$, when $T(t)$ vanishes. The latter correspond to  kinematical singularities that would also be present in the case of equal mass scattering (in this case at $t=4\mu^2$, where $\mu$ is the mass of the particle, since $z_t=1+2s/(t-4\mu^2)$, and thus $T(t)=t-4\mu^2$) and can be seen to be harmless by noting that from the Froissart-Gribov projection~\eqref{eq:FGformula}, the partial wave coefficient $f_J(t)$ has a zero of order $J$ precisely at $t=(m_1 \pm m_2)^2$. To see this, note that when $t \to (m_1 \pm m_2)^2$ the argument of the $Q_J$ function in the Froissart-Gribov projection approaches infinity, and therefore using the asymptotic behaviour $Q_J(z) \sim z^{-J-1}$ we have
\begin{align*}
 f_J(t) &={1\over16 \pi^2}\frac{t}{T(t)}\int\limits_{t_0}^\infty \mathrm{d}t \,Q_J\!\left(1+\frac{st}{T(t)}\right){\rm Disc}_s A(s,t)\\
 &\underset{t\to (m_1 \pm m_2)^2}{\sim} \frac{t}{T(t)}\int\limits_{t_0}^\infty \mathrm{d}t \,\left(\frac{T(t)}{t}\right)^{J+1}s^{-J-1}{\rm Disc}_s A(s,t)=\left(\frac{T(t)}{t}\right)^{J} \times \text{finite}.
\end{align*}

This shows that when combining the contribution from the Legendre function and the partial wave coefficient to the Regge residue at the pole $J=\alpha_0 (t)$, the overall $\left(\frac{t}{T(t)}\right)^{\alpha_0(t)}$ factor in eq.~\eqref{eq:noneqmasses} cancels out, and the factors of $T(t)$ appearing in subleading terms have all positive powers and therefore do not produce poles. This would be the end of the story when all the masses of the scattered states are equal, and thus the Sommerfeld-Watson transform does not have spurious singularities in that case.

However, in the case of non-equal masses, we also find the poles at $t=0$ coming from the fact that $t$ appears both in the denominator and in the numerator in $z_t$. While the leading $1/t^{\alpha_0(t)}$ term is cancelled by the partial wave coefficients, the subleading terms in the asymptotic expansion of the Legendre function add poles at $t=0$ of increasingly high order. 

These higher-order poles are clearly unphysical and need to be cancelled by some mechanism, and the mentioned references argue that using the Mandelstam-Sommerfeld-Watson transform, it is possible to push the contour of the background integral arbitrarily to the left so that it yields a negligible contribution. Subsequently, they postulate the necessity of infinitely many {\it daughter} trajectories, parallel to the leading one and with intercepts shifted by integers, namely $\alpha_n(t)=\alpha_n (t)-i$ for the $n^\text{th}$ trajectory. These are required to remove all of these poles one by one by demanding that $\alpha_n(t)$ has an order $n$ pole at $t=0$ with the appropriate residue to cancel the corresponding term in~\eqref{eq:noneqmasses}. 

However, as we have explained in section~\eqref{sec:mandelstam-regge-poles}, we do not have control over the behaviour of the partial wave coefficients at large, negative $J$, and therefore it is not possible to safely neglect the background integral in general. In fact, as we mentioned in that section, the Veneziano amplitude provides an explicit example of a scenario in which the background integral is not negligible, causing the sum over Regge poles in the Mandelstam-Sommerfeld-Watson transform to be only asymptotic if one does not consider the background integral. Moreover, the case of the Veneziano amplitude even gives an example of spurious singularities being cancelled by the background integral, since the $Q_{-\alpha(t)-1}(-z)/\cos \pi \alpha(t)$ terms in the Regge poles have logarithmic singularities at $z= \pm1$, with branch cuts extending from $z=-\infty$ to $z=1$, and poles at half integer values of $\alpha(t)$, all of which are unphysical and are precisely cancelled by the background integral.

This leads us to observe that it would be completely possible, and actually seems rather generic, that the background integral cancels the aforementioned singularities. This then weakens drastically the claim of the need for daughter trajectories in the case of non-equal masses. Overall, this argument appears invalid and requires assuming extra constraints on the behaviour of the partial wave coefficients which has not been, to our knowledge, justified in the literature.

\subsection{Generalisations}

\paragraph{Situation for Conformal Field Theories}

Due to the higher amount of symmetry contained in the conformal group, the situation for conformal field theories (CFTs) is under better control. In particular, it is known from lightcone bootstrap that in $d\geq2$ dimensions there exist infinitely many trajectories of double-trace operators which approach trajectories of constant twist~\cite{Komargodski:2012ek,Fitzpatrick:2012yx}. This has been proven rigorously recently in~\cite{Pal:2022vqc}, and we refer to the references cited in there for a more thorough discussion of the question in CFTs. Interestingly, the authors of this paper note that the accumulation of such higher spin operators in QFTs was first studied in the context of QCD in~\cite{Parisi:1973xn,Callan:1973pu}.\footnote{We thank Jiaxin Qiao for a discussion on these points.}

\paragraph{Magnitude of daughter trajectories.}

At the end of sec.~\ref{sec:finite-RP}, we explained why our argument based on the regularity of the partial waves fails in the case of infinitely many trajectories. In sec.~\ref{sec:Veneziano}, we saw explicitly that, while each trajectory yields a finite contribution to $A_J(s)$, their infinite sum generates the divergence of $A_J(s)$ at $J=\alpha(s)$. This argument did not constrain the size of the trajectory.

The same argument can be made in the case with subtractions, and in particular, we are led to conjecture that the pion amplitude of~\cite{Albert:2022oes,Albert:2023jtd,Albert:2023seb} should possess infinitely many Regge trajectories, but we cannot give a lower bound on their size with the arguments of this paper.

\paragraph{External spinning particles}
We looked at the specific case of dual model amplitudes with external scalars. What if the external particles now have spin $J_0$? For instance, what about a four-vector boson or four-graviton amplitude? In this case, the spin-$J_0$--spin-$J_0$--spin-$J$ vertex is comprised of more tensorial combinations, and we do not merely decompose on Legendre polynomials but on Wigner $d$-matrices, see e.g.~\cite{Bern:2021ppb}. For gravitons, we could also not keep the colour-ordering assumption and should include explicitly the $u$-channel poles. 

We have not studied this case in detail but we expect that it should be ruled out, too. This expectation derives from the similarity between adding more trajectories and adding tensor structures: any finite number should not allow for the presence of Regge poles in amplitudes with external spinning particles.
In particular, the crucial argument of our proof relies on the convergence of the $s$-channel pole expansion in $t$ the cross channel, which forces the residue coefficients to be exponentially suppressed. Since Wigner $d$-matrices are also hypergeometric functions, we would expect the same kind of exponential growth, forcing a similar exponential decay of the coefficients, which would then, in turn, give rise to regular partial wave coefficients. A more precise argument could be made using the results of~\cite{Balasubramanian:2021act}, where it was shown how to decompose the spinning amplitudes on scalar amplitudes.

\paragraph{Regge cuts, relaxing narrow-width approximation.} 
Finally, it would be very interesting to extend our work to the case of non-pure-meromorphic amplitudes and describe resonances with a non-vanishing width, amplitudes with cuts and Regge cuts, etc. Some remarks had been made in the past in~\cite{atkinson_conditions_1968,Atkinson:1969fe} and it would be interesting to revisit this story within the modern perspective of the $S$-matrix bootstrap. For instance, it is not obvious that a single Regge pole and a Regge cut is a disallowed configuration, but the techniques developed in this paper would not allow us to say anything about this case.

\medskip
\medskip
\medskip

\hrule
\medskip

\noindent\textit{Acknowledgements.}
We would first like to thank Zhenia Skvortsov for asking us the question of whether or not amplitudes with only one trajectory exist. We would also like to thank Johan Henriksson, Yue-Zhou Li, Jiaxin Qiao, Leonardo Rastelli, Slava Rychkov, Aninda Sinha, Alessandro Vichi and Sasha Zhiboedov for discussions and comments. This work has received funding from Agence Nationale de la Recherche (ANR), project ANR-22-CE31-0017. The work of CE is supported by the ANR through grant ANR-19-CE31-0005-01 (PI: F.~Calore), and has been supported by the EOSC Future project which is co-funded by the European Union Horizon Programme call INFRAEOSC-03-2020, Grant Agreement 101017536. This publication is supported by the European Union's Horizon Europe research and innovation programme under the Marie Sk\l odowska-Curie Postdoctoral Fellowship Programme, SMASH co-funded under the grant agreement No.~101081355.

\appendix

\section{Finite Energy Sum Rules}
\label{app:fesr}

In this appendix, we briefly explain the notion of FESRs, to make the discussion on past works in sec.~\ref{sec:discussion}  more self-contained.

FESRs are a consequence of the analyticity of the scattering amplitude in the Mandelstam variables. The idea is as follows: Consider the amplitude as a function of $s$ for fixed $t$ and the following contour integral in the complex $s$-plane:
\begin{equation}
\label{eq:FESR}
\mathcal{I}_k (t,R)=\int_\mathcal{C}\frac{\mathrm{d} s'}{2 \pi i}s'^k A(s',t),
\end{equation}
where $\mathcal{C}$ is a circle of radius $R$. 

This contour integral can be shrunk to the poles inside the circle. If the largest mass in the spectrum such that $m_n^2 < R$ is $m_{N_\text{max}}^2$, we have
\begin{equation}
\label{eq:FESRlhs}
\mathcal{I}_k (t,R)=\sum_{n=0}^{N_\text{max}}(m_n^2)^k P_n(t),
\end{equation}
where $P_n (t)$ is the residue of the amplitude at $s=m_n^2$. 

On the other hand, if $R$ is sufficiently large we might be able to approximate $\mathcal{I}_k (t,R)$ by the integral of the Regge limit of the amplitude along this contour. If we consider only one trajectory we have in this limit
\begin{equation}
\label{eq:reggelimit}
A(s,t)\sim \frac{f(t)e^{- i \pi \alpha(t)}}{\Gamma(\alpha(t)+1) \sin\!{\left(\pi\alpha(t)\right)}} s^{\alpha(t)},
\end{equation}
where $f(t)$ is a residue function solely depending on $t$ (an important fact exploited in some of the past works on single-trajectory dual amplitudes). The integral along the contour $\mathcal{C}$ can be performed explicitly to yield
\begin{equation}
\label{eq:FESRrhs}
\mathcal{I}_k (t,R)=\frac{f(t)R^{\alpha(t)+k+1}}{\pi \Gamma(\alpha(t)+1)(\alpha(t)+1+k)},
\end{equation}
assuming that $R \gg t$, such that corrections to \eqref{eq:FESRrhs} can be neglected.

Two main difficulties emerge in this context. The first one, which is not addressed in any piece of the literature, is the question of the speed of reggeization at arguments of complex $s$ along the integration contour $\cal C$ in~\eqref{eq:FESR}. While it is conceivable that this speed does not depend on the angle, this is an unproven assumption, and our method prevents having to assume anything of the sort. 
Another difficulty is that the integration contour varies discontinuously as it jumps from one pole to the next and picks up a new residue when we increase the radius $R$. However, the approximation in terms of the reggeized amplitude from eq.~\eqref{eq:FESRrhs} is a continuous function of $R$. This incompatibility led previous works to employ smoothing procedures that are plausible but to our understanding not rigorous even though they were fundamental for the previous attempts to study the reggeization of single-trajectory dual amplitudes. 

\bibliographystyle{JHEP}
\bibliography{biblio.bib}

\providecommand{\href}[2]{#2}\begingroup\raggedright\begin{thebibliography}{10}

\bibitem{Paulos:2016fap}
M.~F. Paulos, J.~Penedones, J.~Toledo, B.~C. van Rees and P.~Vieira, \emph{{The S-matrix bootstrap. Part I: QFT in AdS}}, \href{http://dx.doi.org/10.1007/JHEP11(2017)133}{\emph{JHEP} {\bfseries 11} (2017) 133}, [\href{https://arxiv.org/abs/1607.06109}{{\ttfamily 1607.06109}}].

\bibitem{Paulos:2016but}
M.~F. Paulos, J.~Penedones, J.~Toledo, B.~C. van Rees and P.~Vieira, \emph{{The S-matrix bootstrap II: two dimensional amplitudes}}, \href{http://dx.doi.org/10.1007/JHEP11(2017)143}{\emph{JHEP} {\bfseries 11} (2017) 143}, [\href{https://arxiv.org/abs/1607.06110}{{\ttfamily 1607.06110}}].

\bibitem{Paulos:2017fhb}
M.~F. Paulos, J.~Penedones, J.~Toledo, B.~C. van Rees and P.~Vieira, \emph{{The S-matrix bootstrap. Part III: higher dimensional amplitudes}}, \href{http://dx.doi.org/10.1007/JHEP12(2019)040}{\emph{JHEP} {\bfseries 12} (2019) 040}, [\href{https://arxiv.org/abs/1708.06765}{{\ttfamily 1708.06765}}].

\bibitem{Kruczenski:2022lot}
M.~Kruczenski, J.~Penedones and B.~C. van Rees, \emph{{Snowmass White Paper: S-matrix Bootstrap}},  \href{https://arxiv.org/abs/2203.02421}{{\ttfamily 2203.02421}}.

\bibitem{Caron-Huot:2016icg}
S.~Caron-Huot, Z.~Komargodski, A.~Sever and A.~Zhiboedov, \emph{{Strings from Massive Higher Spins: The Asymptotic Uniqueness of the Veneziano Amplitude}}, \href{http://dx.doi.org/10.1007/JHEP10(2017)026}{\emph{JHEP} {\bfseries 10} (2017) 026}, [\href{https://arxiv.org/abs/1607.04253}{{\ttfamily 1607.04253}}].

\bibitem{Haring:2023zwu}
K.~H\"aring and A.~Zhiboedov, \emph{{The Stringy S-matrix Bootstrap: Maximal Spin and Superpolynomial Softness}},  \href{https://arxiv.org/abs/2311.13631}{{\ttfamily 2311.13631}}.

\bibitem{Cheung:2023adk}
C.~Cheung and G.~N. Remmen, \emph{{Stringy dynamics from an amplitudes bootstrap}}, \href{http://dx.doi.org/10.1103/PhysRevD.108.026011}{\emph{Phys. Rev. D} {\bfseries 108} (2023) 026011}, [\href{https://arxiv.org/abs/2302.12263}{{\ttfamily 2302.12263}}].

\bibitem{Cheung:2023uwn}
C.~Cheung and G.~N. Remmen, \emph{{Bespoke dual resonance}}, \href{http://dx.doi.org/10.1103/PhysRevD.108.086009}{\emph{Phys. Rev. D} {\bfseries 108} (2023) 086009}, [\href{https://arxiv.org/abs/2308.03833}{{\ttfamily 2308.03833}}].

\bibitem{Geiser:2022exp}
N.~Geiser and L.~W. Lindwasser, \emph{{Generalized Veneziano and Virasoro amplitudes}}, \href{http://dx.doi.org/10.1007/JHEP04(2023)031}{\emph{JHEP} {\bfseries 04} (2023) 031}, [\href{https://arxiv.org/abs/2210.14920}{{\ttfamily 2210.14920}}].

\bibitem{tHooft:1973alw}
G.~'t~Hooft, \emph{{A Planar Diagram Theory for Strong Interactions}}, \href{http://dx.doi.org/10.1016/0550-3213(74)90154-0}{\emph{Nucl. Phys. B} {\bfseries 72} (1974) 461}.

\bibitem{Witten:1979kh}
E.~Witten, \emph{{Baryons in the 1/n Expansion}}, \href{http://dx.doi.org/10.1016/0550-3213(79)90232-3}{\emph{Nucl. Phys. B} {\bfseries 160} (1979) 57--115}.

\bibitem{Albert:2022oes}
J.~Albert and L.~Rastelli, \emph{{Bootstrapping pions at large N}}, \href{http://dx.doi.org/10.1007/JHEP08(2022)151}{\emph{JHEP} {\bfseries 08} (2022) 151}, [\href{https://arxiv.org/abs/2203.11950}{{\ttfamily 2203.11950}}].

\bibitem{Albert:2023jtd}
J.~Albert and L.~Rastelli, \emph{{Bootstrapping Pions at Large $N$. Part II: Background Gauge Fields and the Chiral Anomaly}},  \href{https://arxiv.org/abs/2307.01246}{{\ttfamily 2307.01246}}.

\bibitem{Albert:2023seb}
J.~Albert, J.~Henriksson, L.~Rastelli and A.~Vichi, \emph{{Bootstrapping mesons at large $N$: Regge trajectory from spin-two maximization}},  \href{https://arxiv.org/abs/2312.15013}{{\ttfamily 2312.15013}}.

\bibitem{Fernandez:2022kzi}
C.~Fernandez, A.~Pomarol, F.~Riva and F.~Sciotti, \emph{{Cornering large-N$_{c}$ QCD with positivity bounds}}, \href{http://dx.doi.org/10.1007/JHEP06(2023)094}{\emph{JHEP} {\bfseries 06} (2023) 094}, [\href{https://arxiv.org/abs/2211.12488}{{\ttfamily 2211.12488}}].

\bibitem{Ma:2023vgc}
T.~Ma, A.~Pomarol and F.~Sciotti, \emph{{Bootstrapping the chiral anomaly at large N$_{c}$}}, \href{http://dx.doi.org/10.1007/JHEP11(2023)176}{\emph{JHEP} {\bfseries 11} (2023) 176}, [\href{https://arxiv.org/abs/2307.04729}{{\ttfamily 2307.04729}}].

\bibitem{Guerrieri:2023qbg}
A.~L. Guerrieri, A.~Hebbar and B.~C. van Rees, \emph{{Constraining Glueball Couplings}},  \href{https://arxiv.org/abs/2312.00127}{{\ttfamily 2312.00127}}.

\bibitem{Guerrieri:2021ivu}
A.~Guerrieri, J.~Penedones and P.~Vieira, \emph{{Where Is String Theory in the Space of Scattering Amplitudes?}}, \href{http://dx.doi.org/10.1103/PhysRevLett.127.081601}{\emph{Phys. Rev. Lett.} {\bfseries 127} (2021) 081601}, [\href{https://arxiv.org/abs/2102.02847}{{\ttfamily 2102.02847}}].

\bibitem{Guerrieri:2022sod}
A.~Guerrieri, H.~Murali, J.~Penedones and P.~Vieira, \emph{{Where is M-theory in the space of scattering amplitudes?}}, \href{http://dx.doi.org/10.1007/JHEP06(2023)064}{\emph{JHEP} {\bfseries 06} (2023) 064}, [\href{https://arxiv.org/abs/2212.00151}{{\ttfamily 2212.00151}}].

\bibitem{Acanfora:2023axz}
F.~Acanfora, A.~Guerrieri, K.~H\"aring and D.~Karateev, \emph{{Bounds on scattering of neutral Goldstones}},  \href{https://arxiv.org/abs/2310.06027}{{\ttfamily 2310.06027}}.

\bibitem{Gumus:2023xbs}
M.~A. Gumus, \emph{{Mapping out EFTs with analytic S-matrix}}, Ph.D. thesis, SISSA, 2023.

\bibitem{EliasMiro:2022xaa}
J.~Elias~Miro, A.~Guerrieri and M.~A. Gumus, \emph{{Bridging positivity and S-matrix bootstrap bounds}}, \href{http://dx.doi.org/10.1007/JHEP05(2023)001}{\emph{JHEP} {\bfseries 05} (2023) 001}, [\href{https://arxiv.org/abs/2210.01502}{{\ttfamily 2210.01502}}].

\bibitem{Eckner:2024ggx}
C.~Eckner, F.~Figueroa and P.~Tourkine, \emph{{The Regge bootstrap, from linear to non-linear trajectories}},  \href{https://arxiv.org/abs/2401.08736}{{\ttfamily 2401.08736}}.

\bibitem{Mandelstam:1963iyb}
S.~Mandelstam, \emph{{Regge Poles and Strip Approximation}},  in \emph{{Theoretical Physics}}, (Vienna), pp.~401--420, IAEA, 1963.

\bibitem{Freedman:1966zec}
D.~Z. Freedman and J.-M. Wang, \emph{{Regge Poles in Unequal-Mass Scattering Processes}}, \href{http://dx.doi.org/10.1103/PhysRevLett.17.569}{\emph{Phys. Rev. Lett.} {\bfseries 17} (1966) 569--572}.

\bibitem{Nachtmann:2003ik}
O.~Nachtmann, \emph{{Pomeron physics and QCD}},  in \emph{{Ringberg Workshop on New Trends in HERA Physics 2003}}, pp.~253--267, 2004, \href{https://arxiv.org/abs/hep-ph/0312279}{{\ttfamily hep-ph/0312279}}, \href{http://dx.doi.org/10.1142/9789812702722_0023}{DOI}.

\bibitem{Arkani-Hamed:2020blm}
N.~Arkani-Hamed, T.-C. Huang and Y.-t. Huang, \emph{{The EFT-Hedron}}, \href{http://dx.doi.org/10.1007/JHEP05(2021)259}{\emph{JHEP} {\bfseries 05} (2021) 259}, [\href{https://arxiv.org/abs/2012.15849}{{\ttfamily 2012.15849}}].

\bibitem{Dolen:1967jr}
R.~Dolen, D.~Horn and C.~Schmid, \emph{{Finite energy sum rules and their application to pi N charge exchange}}, \href{http://dx.doi.org/10.1103/PhysRev.166.1768}{\emph{Phys. Rev.} {\bfseries 166} (1968) 1768--1781}.

\bibitem{Collins_1977}
P.~D.~B. Collins, \emph{An Introduction to Regge Theory and High Energy Physics}.
\newblock Cambridge Monographs on Mathematical Physics. Cambridge University Press, 1977.

\bibitem{Gribov:2003nw}
V.~N. Gribov, \emph{{The theory of complex angular momenta: Gribov lectures on theoretical physics}}.
\newblock Cambridge Monographs on Mathematical Physics. Cambridge University Press, 6, 2007, \href{http://dx.doi.org/10.1017/CBO9780511534959}{10.1017/CBO9780511534959}.

\bibitem{Correia:2020xtr}
M.~Correia, A.~Sever and A.~Zhiboedov, \emph{{An analytical toolkit for the S-matrix bootstrap}}, \href{http://dx.doi.org/10.1007/JHEP03(2021)013}{\emph{JHEP} {\bfseries 03} (2021) 013}, [\href{https://arxiv.org/abs/2006.08221}{{\ttfamily 2006.08221}}].

\bibitem{MANDELSTAM1962254}
S.~Mandelstam, \emph{An extension of the regge formula}, \href{http://dx.doi.org/doi.org/10.1016/0003-4916(62)90218-X}{\emph{Annals of Physics} {\bfseries 19} (1962) 254--261}.

\bibitem{atkinson_conditions_1968}
D.~Atkinson, K.~Dietz and J.~Honerkamp, \emph{Conditions that must be satisfied by crossing-symmetric finite-energy sum-rules}, \href{http://dx.doi.org/10.1007/BF01392968}{\emph{Zeitschrift für Physik} {\bfseries 216} (June, 1968) 281--292}.

\bibitem{Nayak:2017qru}
P.~Nayak, R.~R. Poojary and R.~M. Soni, \emph{{A Note on S-Matrix Bootstrap for Amplitudes with Linear Spectrum}},  \href{https://arxiv.org/abs/1707.08135}{{\ttfamily 1707.08135}}.

\bibitem{Rattazzi:2008pe}
R.~Rattazzi, V.~S. Rychkov, E.~Tonni and A.~Vichi, \emph{{Bounding scalar operator dimensions in 4D CFT}}, \href{http://dx.doi.org/10.1088/1126-6708/2008/12/031}{\emph{JHEP} {\bfseries 12} (2008) 031}, [\href{https://arxiv.org/abs/0807.0004}{{\ttfamily 0807.0004}}].

\bibitem{mandula-slansky-singletraj}
J.~E. Mandula and R.~C. Slansky, \emph{Misuses of the finite-energy sum rules}, \href{http://dx.doi.org/10.1103/PhysRevLett.20.1402}{\emph{Phys. Rev. Lett.} {\bfseries 20} (Jun, 1968) 1402--1405}.

\bibitem{Ademollo:1968cno}
M.~Ademollo, H.~R. Rubinstein, G.~Veneziano and M.~A. Virasoro, \emph{{Bootstrap of meson trajectories from superconvergence}}, \href{http://dx.doi.org/10.1103/PhysRev.176.1904}{\emph{Phys. Rev.} {\bfseries 176} (1968) 1904--1925}.

\bibitem{Fujisaki:1970ce}
H.~Fujisaki, \emph{{Rising regge trajectories and finite energy sum rules}}, \href{http://dx.doi.org/10.1143/PTP.43.101}{\emph{Prog. Theor. Phys.} {\bfseries 43} (1970) 101--113}.

\bibitem{PhysRevLett.21.383}
C.~Goebel, \emph{$t$-channel regge amplitude from $s$-channel resonances}, \href{http://dx.doi.org/10.1103/PhysRevLett.21.383}{\emph{Phys. Rev. Lett.} {\bfseries 21} (Aug, 1968) 383--384}.

\bibitem{Komargodski:2012ek}
Z.~Komargodski and A.~Zhiboedov, \emph{{Convexity and Liberation at Large Spin}}, \href{http://dx.doi.org/10.1007/JHEP11(2013)140}{\emph{JHEP} {\bfseries 11} (2013) 140}, [\href{https://arxiv.org/abs/1212.4103}{{\ttfamily 1212.4103}}].

\bibitem{Fitzpatrick:2012yx}
A.~L. Fitzpatrick, J.~Kaplan, D.~Poland and D.~Simmons-Duffin, \emph{{The Analytic Bootstrap and AdS Superhorizon Locality}}, \href{http://dx.doi.org/10.1007/JHEP12(2013)004}{\emph{JHEP} {\bfseries 12} (2013) 004}, [\href{https://arxiv.org/abs/1212.3616}{{\ttfamily 1212.3616}}].

\bibitem{Pal:2022vqc}
S.~Pal, J.~Qiao and S.~Rychkov, \emph{{Twist Accumulation in Conformal Field Theory: A Rigorous Approach to the Lightcone Bootstrap}}, \href{http://dx.doi.org/10.1007/s00220-023-04767-w}{\emph{Commun. Math. Phys.} {\bfseries 402} (2023) 2169--2214}, [\href{https://arxiv.org/abs/2212.04893}{{\ttfamily 2212.04893}}].

\bibitem{Parisi:1973xn}
G.~Parisi, \emph{{How to measure the dimension of the parton field}}, \href{http://dx.doi.org/10.1016/0550-3213(73)90666-4}{\emph{Nucl. Phys. B} {\bfseries 59} (1973) 641--646}.

\bibitem{Callan:1973pu}
C.~G. Callan, Jr. and D.~J. Gross, \emph{{Bjorken scaling in quantum field theory}}, \href{http://dx.doi.org/10.1103/PhysRevD.8.4383}{\emph{Phys. Rev. D} {\bfseries 8} (1973) 4383--4394}.

\bibitem{Bern:2021ppb}
Z.~Bern, D.~Kosmopoulos and A.~Zhiboedov, \emph{{Gravitational effective field theory islands, low-spin dominance, and the four-graviton amplitude}}, \href{http://dx.doi.org/10.1088/1751-8121/ac0e51}{\emph{J. Phys. A} {\bfseries 54} (2021) 344002}, [\href{https://arxiv.org/abs/2103.12728}{{\ttfamily 2103.12728}}].

\bibitem{Balasubramanian:2021act}
M.~K.~N. Balasubramanian, R.~Patil and A.~Rudra, \emph{{Spinning amplitudes from scalar amplitudes}}, \href{http://dx.doi.org/10.1007/JHEP11(2021)151}{\emph{JHEP} {\bfseries 11} (2021) 151}, [\href{https://arxiv.org/abs/2106.05301}{{\ttfamily 2106.05301}}].

\bibitem{Atkinson:1969fe}
D.~Atkinson and K.~Dietz, \emph{{Infinitely rising regge trajectories and crossing symmetry}}, \href{http://dx.doi.org/10.1103/PhysRev.177.2579}{\emph{Phys. Rev.} {\bfseries 177} (1969) 2579--2581}.

\end{thebibliography}\endgroup

\end{document}